\documentclass{aa}  
\usepackage{natbib}
\newcommand{\starf}{{\it StarFinder}}
\usepackage{graphicx}  
\usepackage{txfonts}
\begin{document}

   \title{The Milky Way's nuclear star cluster: Old, metal-rich, and cuspy}

   \subtitle{Structure and star formation history from deep imaging}

  \author{R. Sch\"odel
          \inst{1}
          \and
          F. Nogueras-Lara
          \inst{2}
          \and
           E. Gallego-Cano
          \inst{3}
          \and
           B. Shahzamanian
          \inst{1}
          \and
          A. T. Gallego-Calvente
          \inst{1}
          \and
          A. Gardini
          \inst{1}
          }

   \institute{
    Instituto de Astrof\'isica de Andaluc\'ia (CSIC),
     Glorieta de la Astronom\'ia s/n, 18008 Granada, Spain
              \email{rainer@iaa.es}
    \and
     Max-Planck Institute for Astronomy, Heidelberg, Germany.
    \and
    Centro Astron\'omico Hispano-Alem\'an, Compl. Observatorio Astron\'omico Calar Alto, s/n
Sierra de los Filabres, 04550 Gergal (Almería), Spain
     }

   \date{;}

% \abstract{}{}{}{}{} 
% 5 {} token are mandatory
 
  \abstract
  % context heading (optional)
  % {} leave it empty if necessary  
  {The environment of Sagittarius\,A* (Sgr\,A*), the central black
    hole of the Milky Way, is the only place in the Universe where we
    can currently study the interaction between a nuclear star cluster
    and a massive black hole and infer the properties of a nuclear
    cluster from observations of individual stars.}
  % aims heading (mandatory)
   {This work aims to explore the star formation history of the nuclear
     cluster and the  structure
     of the innermost stellar cusp around Sgr\,A*.}
  % methods heading (mandatory)
   {We combined and analysed multi epoch high quality AO
     observations. For the region close to Sgr\,A* we apply the speckle
     holography technique to the AO data and  obtain images that are
   $\geq50\%$ complete down to  $K_{s}\approx19$ within a projected
   radius of $5"$ around Sgr\,A*. We used $H$-band images to derive
   extinction maps.}
  % results heading (mandatory)
 {We provide $K_{s}$ photometry for roughly $39,000$ stars and
   $H$-band photometry for
   $\sim$$11,000$ stars within a field of about
     $40"\times40"$, centred on Sgr\,A*. In addition, we provide
   $K_{s}$ photometry of
   $\sim$$3,000$ stars in a very deep central field of $10"\times10"$,
   centred on Sgr\,A*. We find that the $K_{s}$ luminosity function
   (KLF) is rather homogeneous within the studied field and does not
   show any significant changes as a function of distance from the
   central black hole on scales of a few 0.1\,pc. By fitting theoretical
   luminosity functions to the KLF, we derive the star formation
   history of the nuclear star cluster. We find that about 80\%
     of the original star formation took place 10 Gyr ago or longer,
     followed by a largely quiescent phase that lasted for more than
     5\,Gyr. We clearly detect the presence of intermediate-age stars
     of about 3\,Gyr in age. This event makes up about 15\% of the
     originally formed stellar mass of the cluster. A few percent of
     the stellar mass formed in the past few 100\,Myr. Our results
     appear to be inconsistent with a quasi-continuous star formation
     history. The mean metallicity of the stars is consistent with
     being slightly super solar. The stellar density increases
     exponentially towards Sgr\,A* at all magnitudes between
     $K_{s}=15-19$. We also show that the precise properties of the
     stellar cusp around Sgr\,A* are hard to determine because the
     star formation history suggests that the star counts can be
     significantly contaminated, at all magnitudes, by stars that are
     too young to be dynamically relaxed. We find that the probability
     of observing any young (non-millisecond) pulsar in a tight orbit
     around Sgr\,A* and beamed towards Earth is very low. We argue that
     typical globular clusters, such as they are observed in and around
     the Milky Way today, have probably not contributed to the nuclear
     cluster's mass in any significant way. The nuclear cluster may have
     formed following major merger events in the early history of the
     Milky Way. }
  % conclusions heading (optional), leave it empty if necessary 
   {}


   \maketitle
%
%-------------------------------------------------------------------

\section{Introduction}

The vast majority of galactic nuclei of intermediate and high mass galaxies
contain (super)massive black holes (MBHs) and nuclear star clusters
(NSCs). The existence of NSCs is not clear yet in very low mass
galaxies, particularly bulge-less or irregular ones. They are clearly
absent in giant ellipticals \citep[see][]{Neumayer:2012fk,Neumayer:2017fk,Seth:2019zf}. The centre of
the Milky Way contains a massive black hole of
$4\times10^{6}$\,$M_{\odot}$ that is surrounded by a nuclear star
cluster of $2.5\times10^{7}$\,$M_{\odot}$
\citep[][]{Gravity-Collaboration:2018qd,Do:2019ha,Launhardt:2002nx,Schodel:2014fk,Feldmeier-Krause:2017rt}. Since
the Galactic centre (GC) is the nearest galaxy nucleus, with its
distance determined with great accuracy \citep[$8.07\pm0.15$ from
the mean of the values reported in][]{Do:2019ha,Abuter:2019fk}, it is a unique
target in which we can study the properties of a MBH and a NSC as well as
the interaction between them \citep[for an overview, see the reviews
by][]{Genzel:2010fk,Schodel:2014bn}. At present, two intensely studied
questions are the origin of the NSC and the formation of a stellar
cusp by the interaction between the stars and the MBH.

There are two main scenarios that are proposed for the origin of NSCs:
(1) inspiral of clusters and their subsequent merger and (2) star formation 
  in situ. Both processes may contribute with different weights in
different galaxies, and they may occur repeatedly \citep[see][]{Boker:2010ys,Neumayer:2017fk,Seth:2019zf}. The presence of
young stellar populations in the Milky Way's NSC \citep[][and
references therein]{Genzel:2010fk} and in some external NSCs
\citep{Seth:2006uq} provides evidence for the in situ scenario. The
cluster infall scenario is harder to test observationally
because of the long timescales and because infall is expected to have
occurred billions of years in the past. NSC formation via the merger of
recently formed super star clusters might be ongoing in the system
Henize2-10 \citep{Nguyen:2014uq}.

The star formation history of the Milky Way's NSC (MWNSC) can provide
us with clues to its origin. In situ formation is being observed in
the MWNSC, but cluster infall may have contributed in the past, too
\citep{Feldmeier:2014kx}. Of the order of 80\% or more of the stellar
mass appears to have formed more than 5\,Gyr ago, but there is also
evidence for intermediate age stars
\citep{Blum:2003fk,Pfuhl:2011uq}. The mean metallicity of the MWNSC
appears to be solar to supersolar
\citep{Do:2015ve,Feldmeier-Krause:2017kq,Rich:2017rm,Nandakumar:2018zr,Schultheis:2019lw}
and there is a dearth of RR Lyrae stars in the central parsecs
\citep{Dong:2017zl}. These observations suggest a relatively high
metallicity of the MWNSC, which weighs against globular clusters
having contributed any significant amount of stellar mass to
it.

The old age of most of the MWNSC's stars means that a large fraction
of them are probably dynamically relaxed. In this case, theory
predicts the formation of a stellar cusp around the central MBH
\citep[see review by][and references therein]{Alexander:2017fk}. The
observational signature of a stellar cusp is a power-law increase of
the stellar density towards the black hole,
$\rho(r)\propto r^{\gamma}$, where $r$ is the distance to the MBH and
$\gamma=-1.5$ for the lightest stellar components, typically
$\sim$solar mass stars.  Stellar mass black holes can be considered
heavy particles. If they have had sufficient time to relax
dynamically, then they will settle into a steeper distribution, with
$\gamma_{\mathrm remnants}\approx-2$.  The Galactic centre (GC) is currently the
only place in the Universe where we can test the existence of a
stellar cusp via direct measurements of the stellar surface density.

The stellar cusp at the GC has been at the focus of intense and
partially controversial studies
over the past decade
\citep{Genzel:2003it,Schodel:2007tw,Buchholz:2009fk,Do:2009tg,Bartko:2010fk}
because the presence of young, massive stars, high interstellar
extinction that varies on arcsecond scales, extreme source crowding,
and the unknown contamination by young or intermediate age, non-relaxed
stars are issues that are difficult to tackle. Recent
photometric and spectroscopic work, combined with simulations, appears
to provide, finally, robust evidence for the existence of the
predicted stellar cusp
\citep{Gallego-Cano:2018nx,Schodel:2018db,Baumgardt:2018ad,Habibi:2019tl}.
Recently, evidence for a cusp of stellar mass black holes at the GC was
found from X-ray observations \citep{Hailey:2018eq}.

Here we aim to improve our knowledge of the star formation history
of the MWNSC and on the existence of the stellar cusp around
Sgr\,A*. For this purpose we combine improved image reduction and
analysis procedures with the stacking of high quality archival
adaptive optics imaging data of the central parsec of the GC. We
obtain deeper and more complete star counts than what was previously
available. We can use, for the first time, the luminosity function of
the stars at the GC to obtain constraints on the star formation
history. We confirm the presence of the cusp at somewhat fainter
magnitudes than before, but show also that the contamination by stars
that are not old enough to be dynamically relaxed may be
considerable. Finally, we discuss the implications of our new results.

\section{Data and data reduction}

The main limitations in observations of the GC are extinction,
crowding, and the need for a high dynamic range
\citep[see][]{Schodel:2014bn}. The $K_{s}$ band provides high
sensitivity, while minimising extinction in the near-infrared. High
angular resolution adaptive optics observations are required by the
extreme source crowding. Finally, an accurate knowledge of the point
spread function (PSF) is necessary for optimal source detection. The
presence of about a dozen bright stars with magnitudes $K_{s}=6.5-10$,
that can have bright, extended halos in AO observations can make this
task difficult. In order to obtain a high quality and high dynamic
range PSF from the observed field saturation, of the bright stars
should be avoided. We therefore use imaging data with short detector
integration times (DITs) of $\sim$1\,s. Although this means that the
images are read-noise limited, this is no strong limitation on the
data because with a DIT$=1$\,s stars as faint as $K_{s}=20$ can still
be detected with a signal-to-noise ratio of about 10 in one hour of
observing time.

\begin{table}
\centering
\caption{Details of the imaging observations used in this
  work.}
\label{Tab:Obs}
\begin{tabular}{llllll}
\hline
\hline
Date$^{\mathrm{a}}$ & $\lambda_{\rm central}$ & $\Delta\lambda$ & N$^{\mathrm{b}}$ & NDIT$^{\mathrm{c}}$ & DIT$^{\mathrm{d}}$\\
 &  [$\mu$m]  &   [$\mu$m] &  & & [s] \\
\hline
09 May 2010 & 1.66 & 0.33 &  4 & 64 & 2 \\
%27 Sep 2010 & 2.18 & 0.35 &  16 & 126 & 1 \\
09 Aug 2012 & 2.18 & 0.35 & 32 & 60 & 1  \\
11 Sep 2012 & 2.18 & 0.35 &32 & 60 & 1  \\
12 Sep 2012 & 2.18 & 0.35 & 32 & 60 & 1  \\
\hline
\end{tabular}
\begin{list}{}{}
\item[$^{\mathrm{a}}$] UTC date of beginning of night.
\item[$^{\mathrm{b}}$] Number of (dithered) exposures
\item[$^{\mathrm{c}}$] Number of integrations that were averaged on-line by the read-out
  electronics
\item[$^{\mathrm{d}}$] Detector integration time. The total integration time of each observation amounts to N$\times$NDIT$\times$DIT.
\end{list}
 \end{table}
 
 We chose to use the ESO NACO/VLT data listed in Table\,\ref{Tab:Obs},
 which are almost identical to the data used in
 \citet{Gallego-Cano:2018nx}.  After standard infrared imaging
 reduction (sky subtraction, flat fielding, bad pixel correction),
 rebinning by a factor of two (cubic interpolation), and alignment of
 the images via the shift-and-add algorithm and a bright reference
 star, we created a deep mosaic of the large field. We applied cubic
 interpolation for rebinning the images. Since the images are barely
 Nyquist sampled, this improves the photometry and astrometry of the
 final products \citep[see
 also][]{Gallego-Cano:2018nx,Schodel:2018db}. An uncertainty map
 was obtained from the errors of the means of the individual pixels.

\begin{figure}[!htb]
\includegraphics[width=\columnwidth]{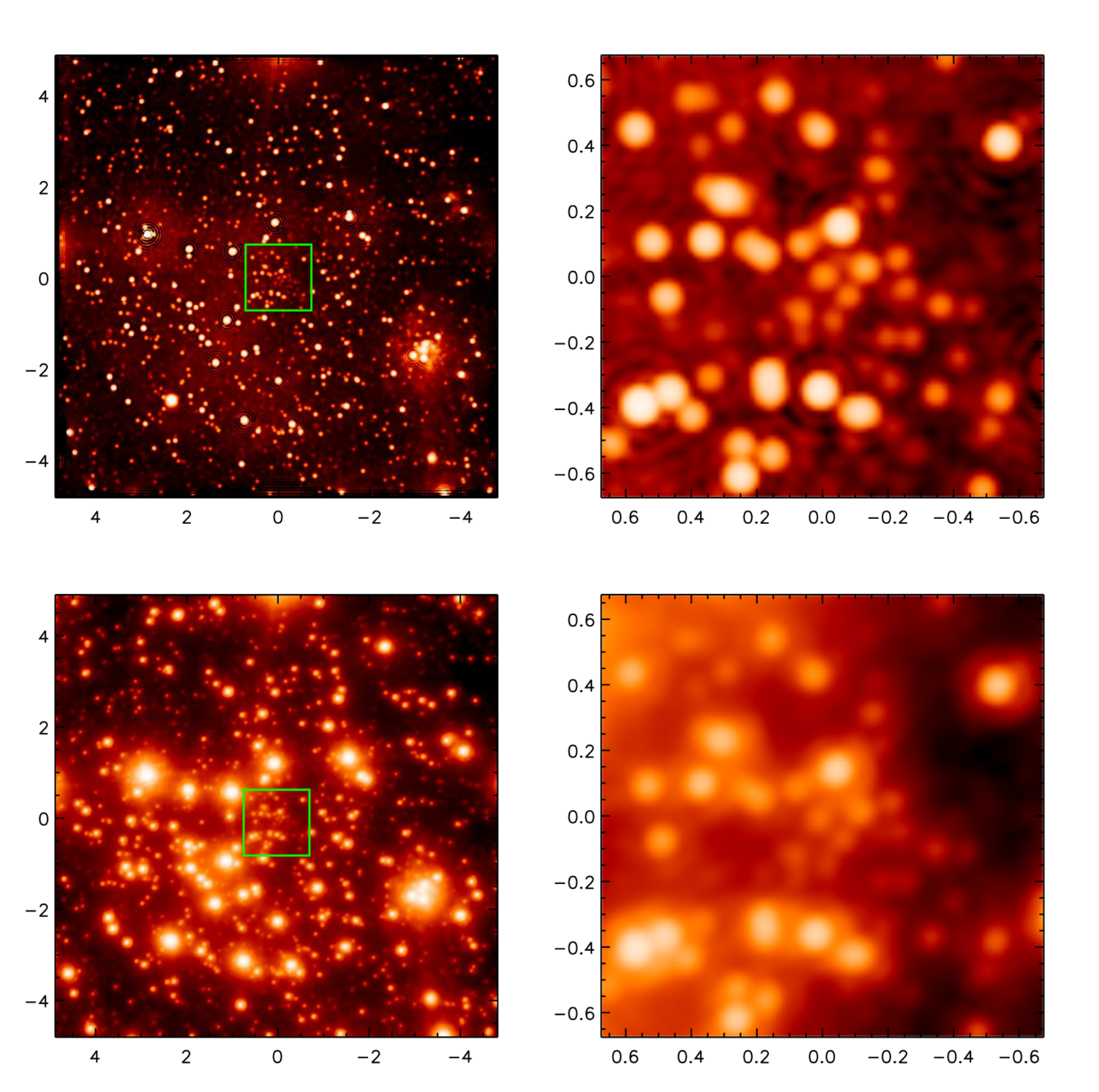}
\caption{\label{Fig:images} Upper left: Deep image after combining all
  2012 $K_{s}$ exposures with the holography technique. Lower left:
  Deep image after combining all 2012 $K_{s}$ exposures with the SSA
  technique. The insets on the right hand side show magnifications of
  the areas marked by the green squares. Colour scales are
  logarithmic. The axes mark offsets from Sgr\,A* along Right
  Ascension and Declination in arcseconds.}
\end{figure}
 
The central, most crowded region was reduced additionally in a more
complex manner. We extracted a field of $9.72"\times9.72"$ from
each exposure, approximately centred on Sgr\,A*.  For the 2012 $K_{s}$
data we thus obtained 96 image cubes containing 60 exposures each.
Instead of combining the individual exposures by a classical
shift-and-add technique we applied the speckle holography technique as
implemented by \citet{Schodel:2013fk} to combine the exposures. About
10 bright stars in the field were used as PSF references for this
procedure. The resulting 96 holography images were stacked to produce
a final, deep image as well as 100 bootstrapped deep images by drawing
randomly with replacement 96 exposures from this set of holography
images. Pixel uncertainty maps were created from the error of the mean
of the individual values of each pixel. The uncertainty maps were used
as noise maps for $\starf$.  For control and benchmarking purposes we
also created a deep image as well as corresponding bootstrapped images
with the simple shift-and-add (SSA) technique. Both the deep image
from the holography procedure, which we refer to as the {\it
  holographic image} in the rest of the text, and the deep SSA image
are shown in Fig.\,\ref{Fig:images}.

\section{Astrometry, photometry, and completeness}

\subsection{Source detection}
\label{sec:detection}

We used the $\starf$  software for photometry and astrometry
\citep{Diolaiti:2000qo}. For the large $K_{s}$ mosaic we chose a
detection threshold of $3\,\sigma$, with two iterations. $\starf$ was
run three times, with three different correlation thresholds of
$0.60, 0.70$, and $0.80$, thus resulting in three different lists that
reflect the uncertainty of the setting of the correlation threshold
value. A spatially variable PSF was used, as described in
\citet{Gallego-Cano:2018nx}. In particular, the PSF was determined for
small $13.8"\times13.8"$ sub-fields. The pattern of sub-fields
overlapped by $6.9"$. The multiple measurements in the overlap regions
were used to estimate the uncertainty introduced by the variable PSF
\citep[see also][]{Schodel:2010hc}. The latter uncertainty term is
approximately constant for all magnitudes and across the mosaic. It
was added in quadrature to the formal uncertainties returned by
{\it StarFinder}.

A different procedure was applied to the central field, which is the
most crowded area and therefore most prone to errors. The central
field is small enough to perform computationally intensive work. We
used the bootstrapped images to infer astrometric and photometric
uncertainties. We proceeded as follows: (1) Run $\starf$ on the deep
image with three iterations, using a $3\,\sigma$ detection threshold,
a minimum correlation coefficient of $0.7$, a $back\_box$ parameter of
eight pixels (it is used to estimate the diffuse background), and
applying deblending of very close sources. $\starf$ returns a
correlation value of $-1.0$ for extended sources. They were excluded
from the list of detected stars.  We thus obtained what we call the
'master list'.  (2) Run $\starf$ on the 100 bootstrapped images with
the same parameters. (3) Compare the master list with the lists
obtained from the bootstrapped images. Sources detected within $0.04"$
of a star in the master list are considered to be the same star. (4)
Create a final list with stars that are detected in $\geq70\%$ of the
bootstrapped images. The astrometric and photometric positions and
uncertainties are computed from the mean and standard deviation of the
measurements in the bootstrapped images,  after having corrected
  for small systematic offsets between the lists from the bootstrapped
  images and from the deep image.

\subsection{Calibration}

Astrometric calibration was achieved using the reported positions and
proper motions of the maser stars IRS\,9, IRS\,7 IRS\,12N, IRS\,28,
IRS\,17, IRS\,10EE, and IRS\,15NE \citep{Reid:2007vn}. The uncertainty
of the astrometric calibration was estimated by comparing the
recovered positions after repeating the alignment procedure, dropping
a different one of the maser sources each time. The star IRS\,17
showed the greatest standard deviation with about $0.01"$. We did not
take any optical distortion of the S27 camera into account
\citep[e.g.][]{Plewa:2015yu}. However, precision (milli-arcsecond)
astrometry is not relevant to the scientific purpose of this
paper. 

The $H-$band data were treated in the same way as the
$K_{s}$-band data and were aligned astrometrically with the
latter. Sources coincident within $0.040"$ after alignment (with a
second order polynomial) were considered to be the same star.
For photometric calibration we used the magnitudes of the stars
IRS\,16C, IRS\,16NW, and IRS\,33N as reported by
\citet{Schodel:2010fk} ($K_{s} = 9.93, 10.14, 11.20$,
$H=11.90,12.03,13.24$).  The resulting uncertainties of the zero points
are $0.08$\,mag. 

Since the field-of-view of the central, $9.72"\times9.72"$ field is
smaller than the anisoplanatic patch in the observed band and since we
did not notice any significant, systematic residuals in the
point-source subtracted images, we did not take into account any
potential uncertainties related to small changes of the PSF across the
field. The analysis and conclusions of this paper do not require any
particularly high precision or accuracy in the photometry and
astrometry.

The astrometric and photometric uncertainties for the holography and
SSA images technique are compared in Fig.\,\ref{Fig:astrophot}. The
astrometric and photometric uncertainties for the holographic and SSA
techniques are similar, but they are somewhat smaller for the
holographic image. The holography procedure
is linear and the large number of individual frames makes sure
that the division in Fourier space, on which the holographic method is
based, does not introduce any significant additional
uncertainty. A significantly larger number of faint sources
is detected in the holographic image (see below). This is probably the case
because the holographic method concentrates the light from the stars
into very compact PSFs, which enhances the contrast of the image,
particularly near the numerous bright stars in the field.

\subsection{Source lists}

\begin{figure}[!htb]
\includegraphics[width=\columnwidth]{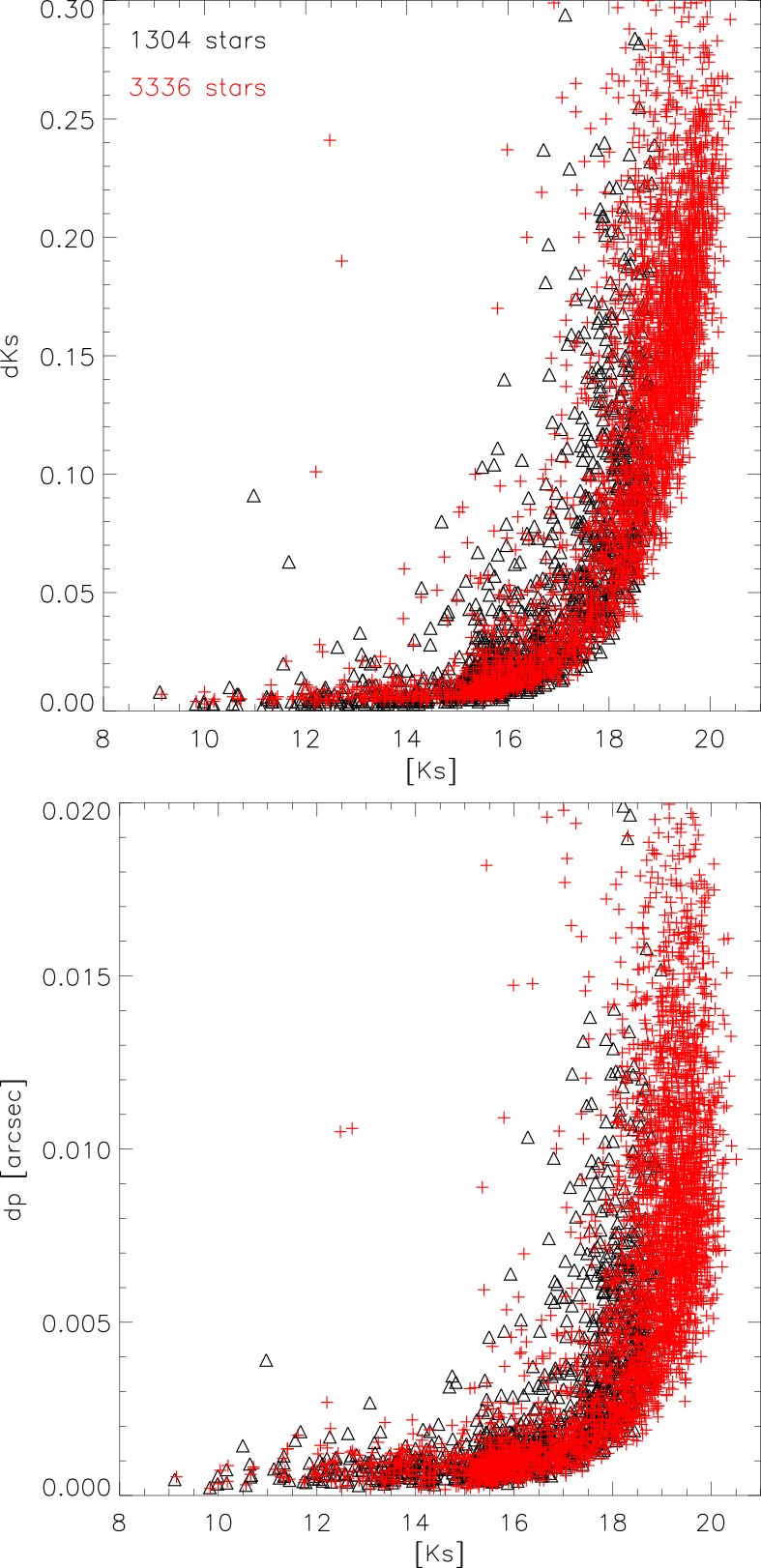}
\caption{\label{Fig:astrophot} Photometric and astrometric
  uncertainties in the central, $9.72"\times9.72"$ field. The upper
  plot shows the $1\,\sigma$ uncertainty of the measured $K_{s}$
  magnitude of the detected stars plotted over the corresponding
  $K_{s}$ magnitude. The lower image shows a plot with the $1\,\sigma$
  relative astrometric uncertainties. Red crosses refer to stars
  measured in the holographic image. Black triangles mark stars measured
  in the SSA image. }
\end{figure}

Table\,\ref{Tab:list_mosaic} lists the positions
and magnitudes of the stars detected in the large field with a
correlation threshold of $0.70$.  There are about $39,000$ detections
at $K_{s}$ with about $11,000$ associated $H$-band measurements.  The
final, full source list is
available electronically.  The 2010 $H$-band star list and the 2012
$K_{s}$-band lists were aligned via a second degree polynomial fit and
stars coincident within $0.04"$ were considered to be the same
source. A proper motion of 1\,mas per year corresponds to about
40\,km\,s$^{-1}$ at the distance of the GC. Therefore, even stars near
Sgr\,A*, moving at several hundreds of km\,s$^{-1}$, will be
cross-identified by this procedure because the time baseline is only
two years.

\begin{figure}[!htb]
\includegraphics[width=\columnwidth]{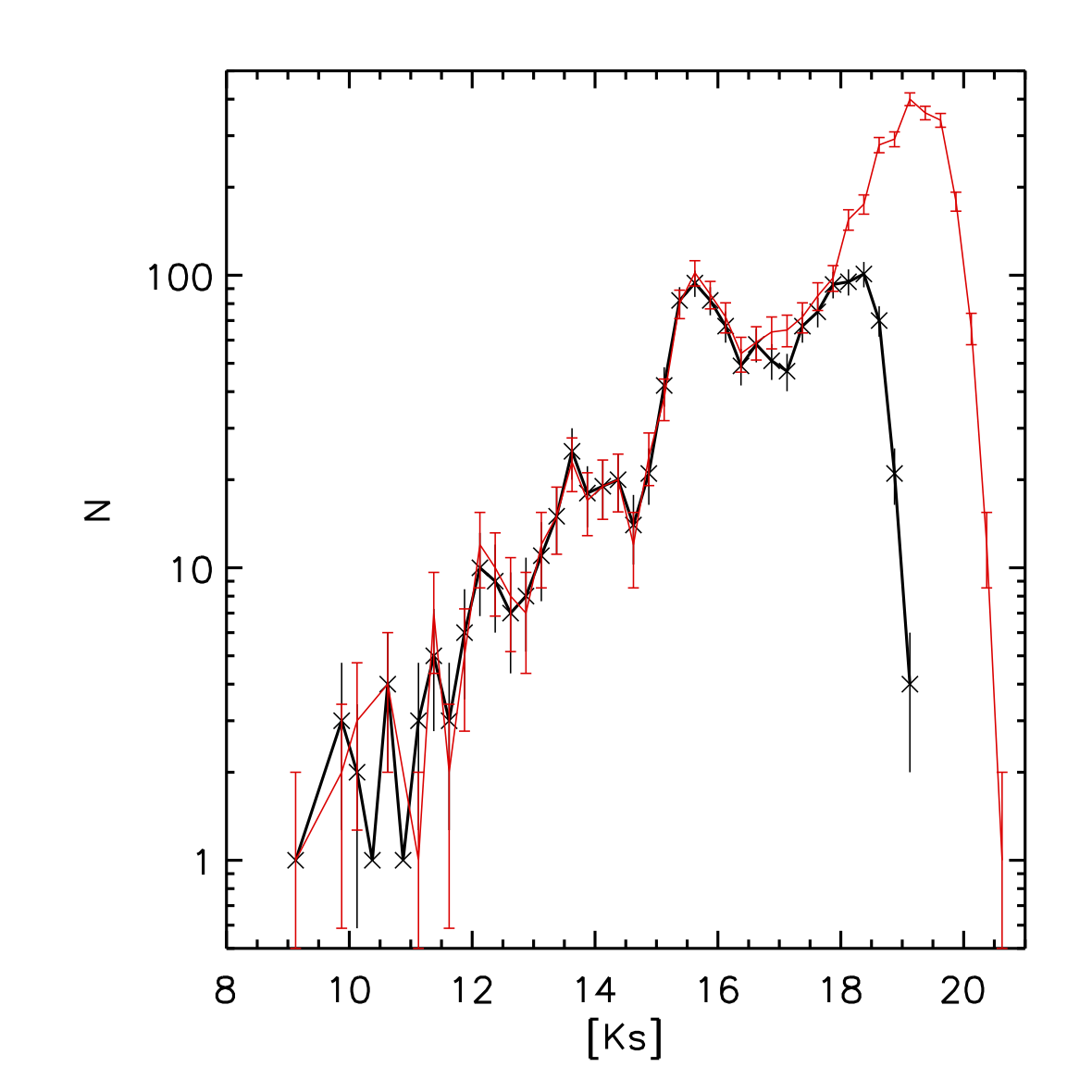}
\caption{\label{Fig:KLF} KLF for the stars
  detected in the holographic image (red) and in the SSA image (black). }
\end{figure}

By following the procedure outlined in the
preceding paragraphs, we identified 3336 stars in the
holographically treated data and 1304 stars in the SSA image for
2012 in the central field. Almost all the additionally detected stars in the image
processed via holography were detected at faint magnitudes. The
astrometric and photometric uncertainties were found to be smaller for
the holographically treated image than for the SSA image
(Fig.\,\ref{Fig:astrophot}).  Figure\,\ref{Fig:KLF} shows a comparison
of the raw $K_{s}$ luminosity functions (KLFs) of the SSA and
holographic image. Table\,\ref{Tab:list_central} lists the positions
of the stars relative to Sgr\,A*, their $K_{s}$-magnitudes and
uncertainties. The full list is available electronically.

\subsection{Completeness}

Completeness is usually estimated via the method of artificially
inserting stars into images. In the GC observations used here,
completeness is mainly limited by stellar crowding. A special  problem in the complicated GC field is the
challenge to avoid including spurious sources into the final star
lists. As already discussed in \citet{Gallego-Cano:2018nx} or
\citet{Schodel:2014bn}, the use of noise maps is very efficient in
suppressing the detection of spurious sources. In addition to a
detection significance of $3\,\sigma$, in the case of the
central field we demand that each accepted source be present in 70\%
of the bootstrapped images. Spurious sources can thus be efficiently
suppressed. Unfortunately, this complex procedure, including many
individual exposures, bootstrapped images, and different selection criteria, means that
a completeness test via artificial stars would be very complex and
extremely time consuming. We therefore resort to an alternative
method, as initially described by \citet{Eisenhauer:1998tg} and later
applied\ by \citet{Harayama:2008ph} or \citet{Schodel:2010fk}, for example.

\begin{figure}[!htb]
\includegraphics[width=\columnwidth]{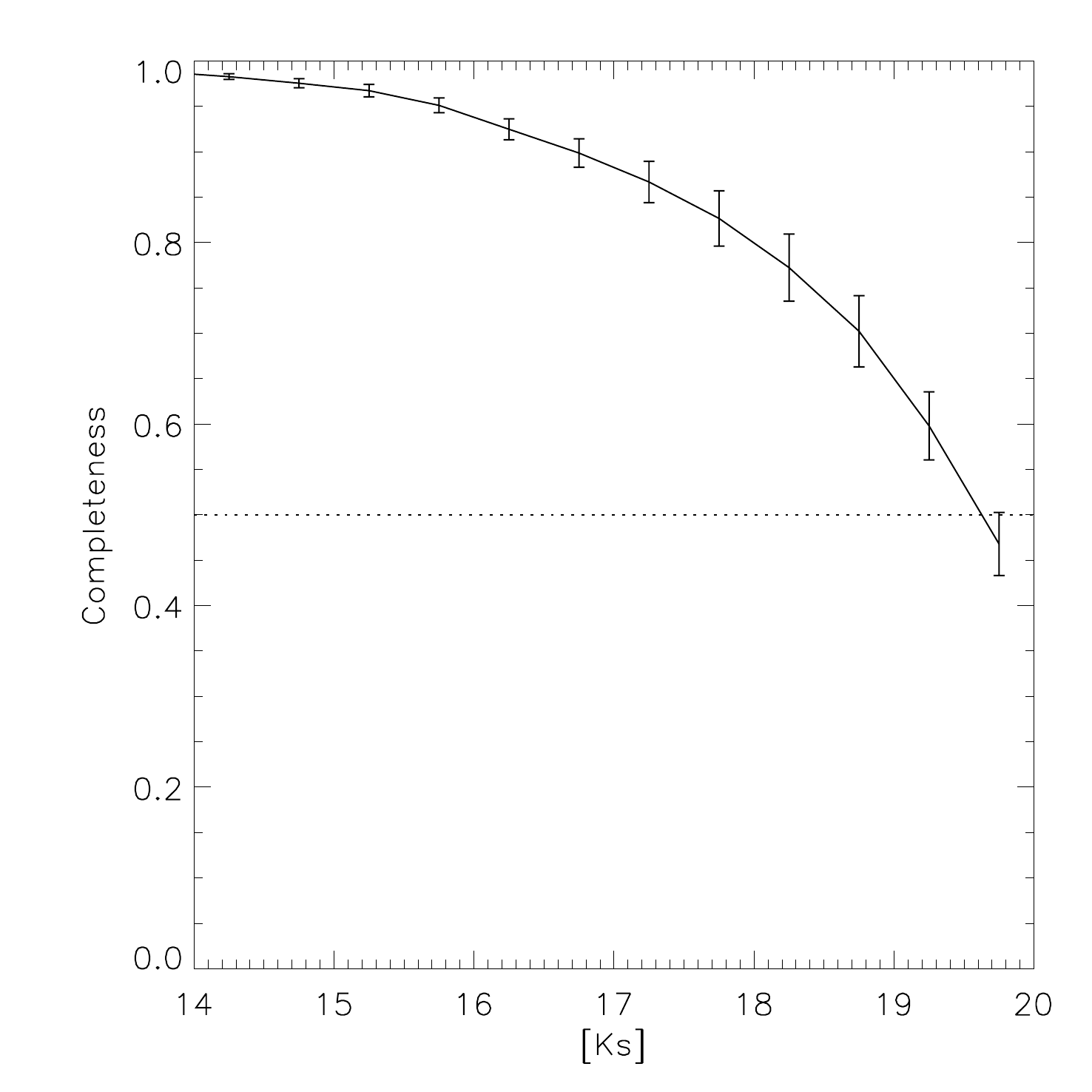}
\caption{\label{Fig:completeness} Completeness as a function of
  stellar magnitude for the central field. The horizontal line
  indicates 50\% completeness due to crowding.}
\end{figure}

In brief, the method is based on an analysis of the distances and
magnitude differences between pairs of detected stars. From these data
the so-called critical distance can be determined for any given
magnitude difference, inside of which the detection probability for a
faint star close to a brighter star drops rapidly. For determining the
critical radius, one assumes a detection probability threshold,
typically 50\%. With the set of magnitude differences and critical
distances one can build completeness maps. The method is based on some
simplifying assumptions, such as a locally approximately constant
density of stars of a given magnitude (Here we mean by ``local'' an
area of the order of one to a few square arcseconds). While being less
direct than the artificial stars technique, the method of
\citet{Eisenhauer:1998tg} has the great advantage that it can be
directly applied to lists of stars that result from complex selection
procedures. Also, it is computationally far less demanding than the
standard method. The inferred completeness fraction for given stellar
magnitudes is shown in Fig.\,\ref{Fig:completeness}. The uncertainties
were estimated by varying the detection probability threshold between
40\% and 60\%.  

We find that the 50\% completeness limit due to crowding is at
$K_{s}\approx19.5$. This is roughly a magnitude deeper than in our
previous work \citep[see Fig.\,3 in][]{Gallego-Cano:2018nx}. The
latter work was based on the same data, but with a final, deep image
obtained via an SSA image. When we analyse the completeness of our SSA
image of the central region with the method used here, we obtain a 50\% completeness limit
of $K_{s}=18.5$.

We also used the above-described procedure to estimate completeness in
the large field. The procedure was applied to the three lists obtained
with three different values of the $\starf$ correlation threshold. We
also varied the probability threshold that is used to infer the
critical radius between 30\% and 70\% to infer the related
uncertainties, which we found to be smaller than 5 percentage points
for the entire considered range of magnitudes. We found that the
completeness due to {\it crowding} was $\gtrsim70\%$ for all sources
$K_{s}\leq18.5$. 

Variable extinction may reduce the completeness of faint stars in
  high-extinction areas. However, deviations from mean extinction in
  the fields observed by us are generally $<0.5$\,mag \citep[see
  extinction maps in][]{Schodel:2010fk}. Since the $3\,\sigma$
  detection limit of our observations is $K_{s}\gtrsim20$, but we
  analyse the luminosity function only down to $K_{s}=19$, we
  therefore consider completeness due to variable extinction not to be
of any concern for this work.

\section{Extinction correction and final source selection}

\begin{figure}[!htb]
\includegraphics[width=\columnwidth]{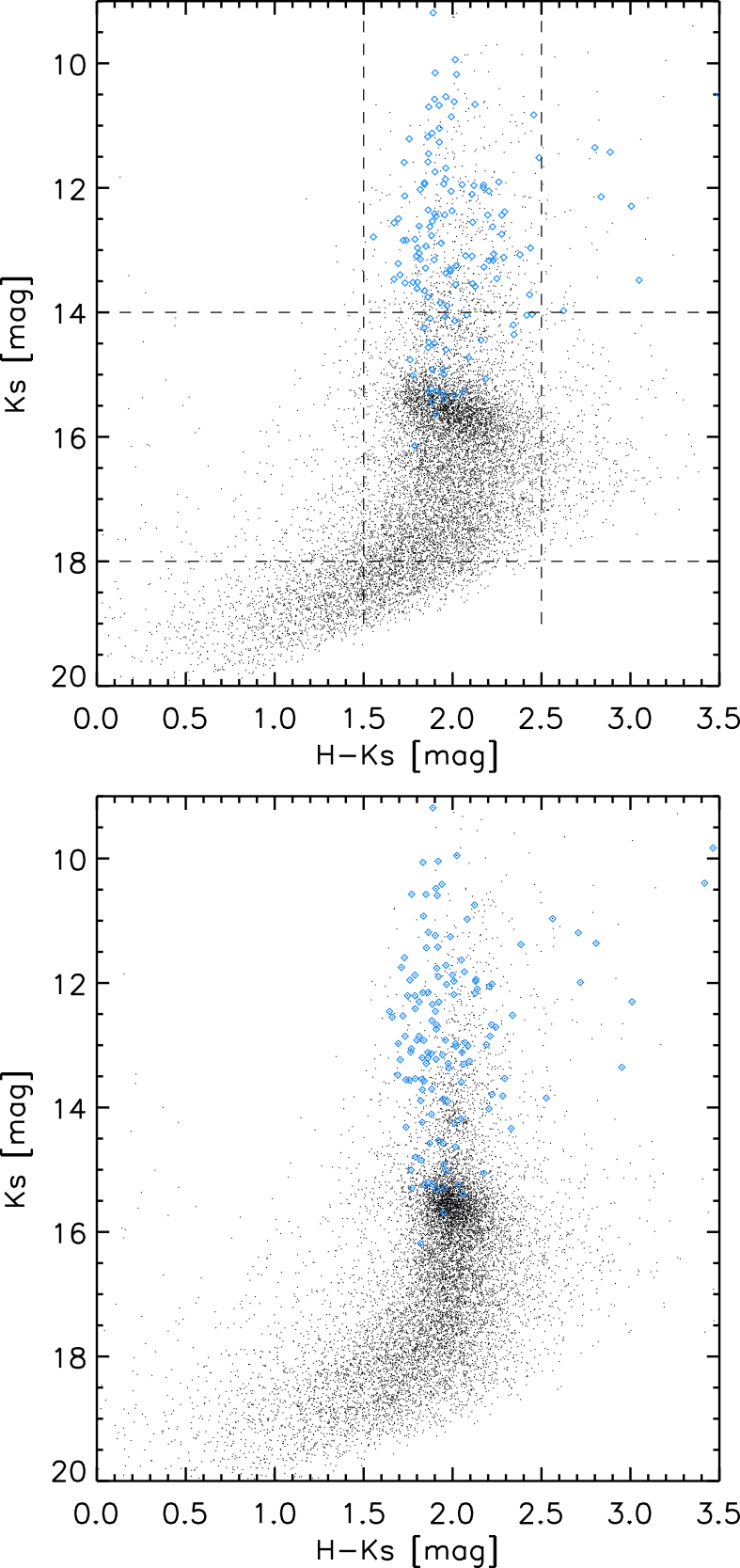}
\caption{\label{Fig:CMDs} Upper panel: $HK_{s}$ colour magnitude
  diagram (CMD) for our data (large field). Colour cuts for the creation of the
  extinction map to exclude foreground stars, highly reddened stars,
  and faint stars, where the $H$-band data become incomplete, are
  indicated by dashed lines. Blue diamonds mark spectroscopically
  identified young, massive stars. Lower panel: Extinction corrected CMD. }
\end{figure}
 
Young, massive stars were identified by cross-matching our source
lists with the spectroscopically identified young stars found by
\citet{Do:2009tg} and \citet{Bartko:2009fq}. For estimating
interstellar extinction,  we excluded foreground
stars and highly (probably intrinsically) reddened sources by red and
blue colour cuts ($H-K_{s}\geq1.4$ and $H-K_{s}\leq2.5$). We additionally limited
the source selection to $18\leq K_{s}\leq14$ as indicated in the
upper panel of Fig.\,\ref{Fig:CMDs}. In this way we include giants of
a mostly homogeneous colour and minimise bias due
to very bright giants and also minimise bias due to the
incompleteness of the $H$-band data for faint stars. 
Subsequently we assumed intrinsic
colours of $H-K{s}=-0.1$ for hot, massive stars and $H-K_{s}=0.1$ for
all other stars. Due to the very limited range of the intrinsic
stellar colours for the chosen filters and potential stellar
population this approach is a well-justified approximation \citep[see,
e.g.\ ][]{Schodel:2010fk}.

We then estimated the extinction for each star individually from the
$H-K_{s}$ colour excess of all neighbouring stars within a projected
radius of $1"$. The mean colour excess was determined with the IDL
astrolib ROBUST\_MEAN procedure in order to exclude any $> 3\,\sigma$
outliers.  Finally, we assumed for the reddening law
$A_{\lambda}\propto\lambda^{-\alpha}$, where $A_{\lambda}$ is the
extinction in magnitude for a given wavelength, $\lambda$. We chose
$\alpha =2.21$ \citep{Schodel:2010fk,Nogueras-Lara:2018pr,Nogueras-Lara:2019zv}. The
typical uncertainties of the derived $A_{K}$ is $0.05$\,mag. The
resulting extinction map is extremely similar to the corresponding part
of the extinction map published by \citet{Schodel:2010fk}. For this
reason we do not show it here. The extinction correction was also
applied to stars that were too faint to have an $H-$band counterpart.

The lower panel of Fig.\,\ref{Fig:CMDs} shows the extinction-corrected
CMD, shifted to the mean extinction of the field. The extinction correction
reduces visibly the scatter of the data points and the Red Clump (RC)
appears like a compact cloud of points centred on $K_{s}\approx15.75$
and $H-K_{s}\approx2.0$. Also, the spectroscopically identified young
stars (blue diamonds), scatter less after the extinction correction
and align approximately along a vertical line to the left of the giant
branch, as is to be expected for hot stars. The
extinction-correction for foreground stars is overestimated, but
they will be excluded from our analysis by the colour cut described in
the first paragraph of this section.

\section{$K_{s}$-Luminosity function}

\subsection{Large field}
\begin{figure}[!htb]
\includegraphics[width=\columnwidth]{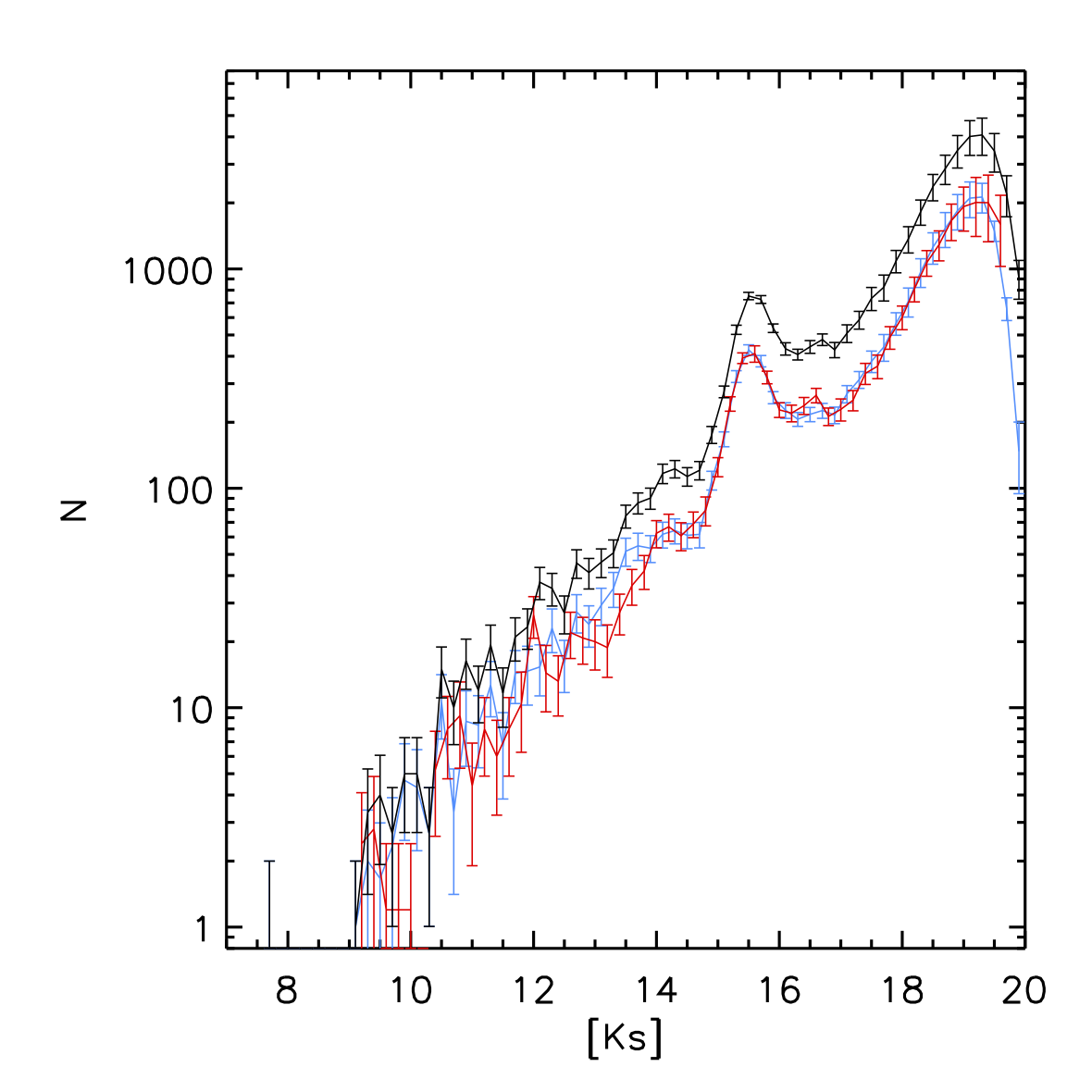}
\caption{\label{Fig:KLF_large} KLF for the large field (black), after
  exclusion of foreground sources,  correction for completeness
    due to crowding, and correction for
    differential extinction, assuming a mean extinction of
    $A_{Ks}=2.62$.  The blue and red lines are the KLFs of the same
  data, but for the region inside (blue) and outside (red) of a
  projected radius of $0.5$\,pc around Sgr\,A*. The red and blue lines
  have been scaled by an arbitrary factor to match at $K_{s}>14$. }
\end{figure}

The extinction and completeness corrected KLF for the large field with
a binning of $0.20$\,mag is shown in Figure\,\ref{Fig:KLF_large}. 
  All stars with colours $H-K_{s}\geq1.4$ were excluded as foreground
  stars \citep[see,e.g.\ ][and discussion therein]{Nogueras-Lara:2018pr,Nogueras-Lara:2019qd}.  The  KLF is the result of averaging the detections from the three $\starf$ runs with
different correlation thresholds (see section\,\ref{sec:detection})
and is shifted to a mean extinction of $A_{K_{s}}=2.62$\,mag. The
uncertainty bars include the Poisson uncertainty (square root of the
counts), the uncertainty of the completeness correction, and the
uncertainty due to using different correlation threshold parameters in
{\it StarFinder} (see Appendix\,\ref{app:KLF_systematics}).

Since the faintest stars have higher photometric uncertainties
($\sim0.3$\,mag at $K_{s}\approx19$), they show a significant scatter
in colour (see Fig.\,\ref{Fig:CMDs}). We considered all stars with
$H-K_{s}<1.4$ (roughly two standard deviations from the mean colour)
as foreground stars.

The blue and red lines in Fig.\,\ref{Fig:KLF_large} show the KLFs for
the regions inside (blue) and outside (red) of a projected radius
$R=0.5$\,pc around Sgr\,A*.  At $K_{s}\leq14$ one can see some
systematic excess in the inner half parsec. It is caused by the roughly 200
spectroscopically identified massive young stars located within a
projected radius of $R=0.5$\,pc around Sgr\,A*. They are strongly
concentrated towards the centre
\citep{Paumard:2006xd,Bartko:2009fq,Lu:2009bl,Yelda:2014fk}. There is
no significant difference between the two KLFs at fainter magnitudes.

\subsection{Central field}
\label{sec:KLF_central}

\begin{figure}[!htb]
\includegraphics[width=\columnwidth]{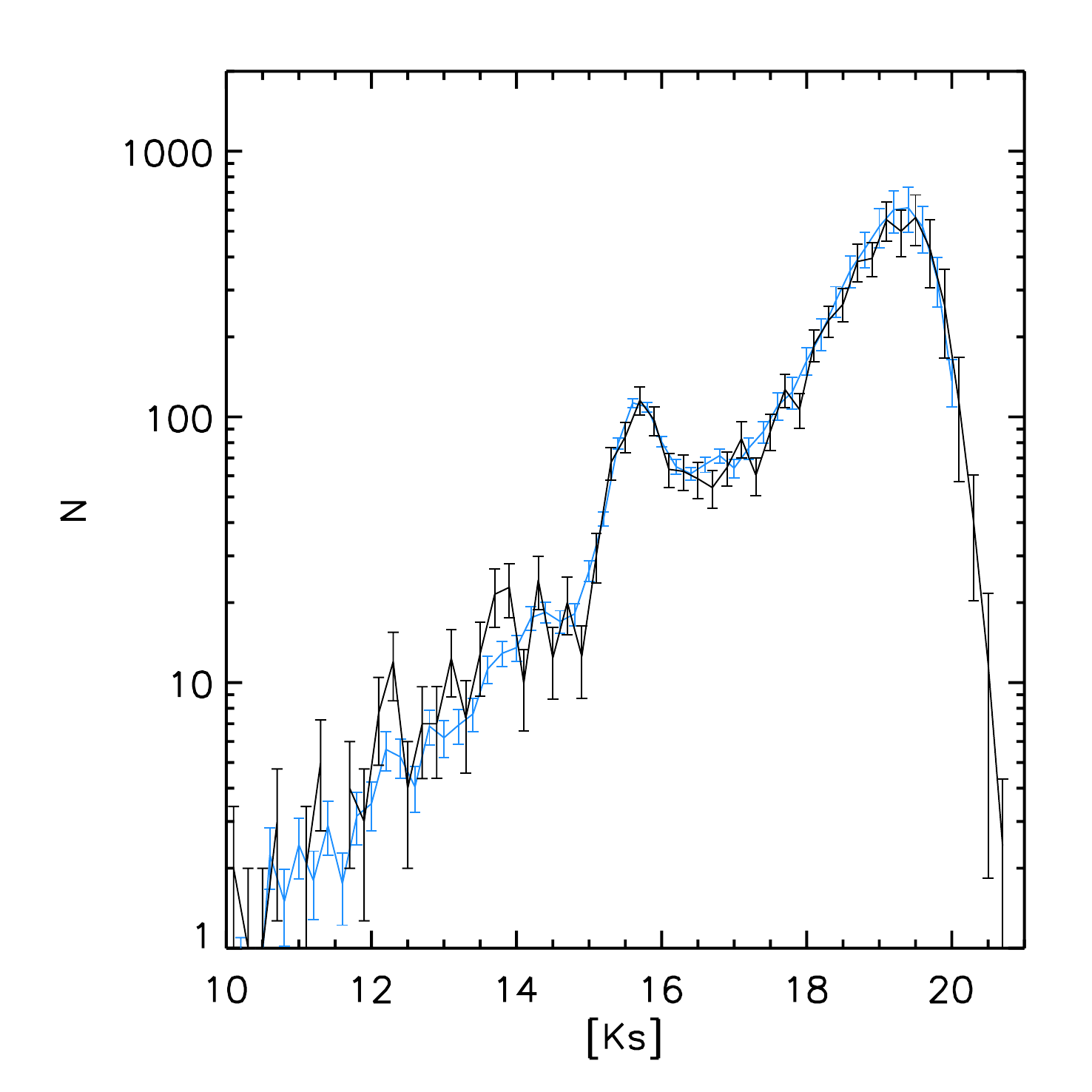}
\caption{\label{Fig:KLF_comp_large_central} Comparison between the KLF
  of the large field (blue, scaled by an arbitrary factor to
  facilitate comparison) and the KLF for the central field
  (black) (both LFs corrected for differential extinction and for 
    completeness due to crowding, with foreground sources excluded).}
\end{figure} 

In Fig.\,\ref{Fig:KLF_comp_large_central} we show a comparison between
the KLF of the large field and the one we created from the star list
of the holographic image of the central field both corrected for
  extinction and crowding, foreground stars excluded. There is no
significant difference, except for the excess due to young massive
stars at $K_{s}\leq14$ in the central field, as mentioned above. We
can see that at the faintest magnitudes the KLF of the central field
does not show any deficiencies. This is in spite of the central field
suffering extreme crowding and the presence of a high concentration of
bright stars, while the number counts of the KLF from the large field
are dominated by much larger and less crowded fields. This vindicates
the application of the holography technique to the central field.

The fact that the KLFs of the central and large regions are
indistinguishable after the exclusion of the spectroscopically
identified young stars supports strongly the notion that the most
recent star formation event near Sgr A* had a top-heavy initial mass
function, as analysed and discussed by, e.g.\ \citet{Bartko:2010fk} and
\citet{Lu:2013fk}. \citet{Gallego-Cano:2018nx} show how the pre-main
sequence stars of this population should show up prominently in the
KLF of the central region if the initial mass function had been of
standard form. The only caveat of this reasoning is that it does not
consider potential mass segregation between the heavier and the
lighter stars of the young population. 

Potential sources of systematics in the creation of the KLF for the
central field are, for example: (1) The acceptance threshold for real
stars in the bootstrap procedure, (2) the colour cut, and (3) the
inclusion of the crowded IRS\,13 area that contains much gas and dust
and may give rise to spurious sources or bias the results because it
may contain young stellar objects \citep[see,
e.g.][]{Muzic:2008fk,Eckart:2013uq}.  Changing the colour cut from
$H-K_{s}=1.4$ to $1.6$ will make the data points vary well within
their $1\sigma$ uncertainties. The same is the case for inclusion or
exclusion of the IRS\,13 area. In order to take the acceptance
threshold into account, we computed the mean KLF and its corresponding
uncertainties resulting from combining acceptance thresholds of 50\%,
70\%, and 90\%. This is the KLF shown in
Fig.\,\ref{Fig:KLF_comp_large_central}.

Since the KLFs of the large and of the central fields are
  indistinguishable within their uncertainties, we only analyse
  the KLF of the large field in the following sections. The KLF of the
large field offers the advantage of smaller statistical uncertainties.

\section{Star formation history}

The KLF can be used to constrain the star formation history of the
MWNSC by comparing it with models created on the basis of theoretical
isochrones. While colour-magnitude diagrams are more powerful tools,
of course, the intrinsic stellar colours in the two near-infrared
bands used in this work are small and the measured colours have
uncertainties $ H-K_{s}\gtrsim0.1$ due to measurement uncertainties,
potential variability (data are not from the same epoch), and the
uncertainty of the extinction correction. Also, since extinction rises
steeply towards shorter wavelengths, completeness at $H$ is
significantly smaller than at $K_{s}$. We therefore use the colour
information only for extinction correction, but not for studying the
stellar population.

We use combinations of single age stellar populations based on
  the updated BaSTI isochrones
  \citep[http://basti.oa-teramo.inaf.it,http://basti-iac.oa-abruzzo.inaf.it/isocs.html,
  ][]{Pietrinferni:2004sp,Bedin:2005th,Cordier:2007fy,Pietrinferni:2013la,Hidalgo:2018mw}.
  The isochrones were computed based on scaled solar metallicities,
  taking overshooting and diffusion into account, assuming a mass loss
  coefficient of $\eta=0.3$ and a Helium fraction of $0.247$. We used
  a Salpeter initial mass function (IMF). Given that the range of
  masses sampled by our observational data is very small (very roughly
  $1-2\,M_{\odot}$, with the exception of the youngest stars) the
  exact IMF used will have very little impact on the results. From the
  BaSTI models we chose isochrones for metallicities of approximately
  twice solar ($[Fe/H]=0.30$\,dex), one and a half times solar ($[Fe/H]=0.15$\,dex), solar ($[Fe/H]=0.06$\,dex) and half solar ($[Fe/H]=-0.30$\,dex).

We limit the fits on the bright end of the KLF to
  $K_{s}\leq 10.0$ because the isochrones used do not extend to
  include brighter stars, and at its faint end to $K_{s}\leq 19.0$
  because of completeness. These magnitudes corresponds to extinction
  corrected magnitudes $7.4\lesssim K_{s}\lesssim16.4$, where
  completeness is above 50\%. We treat the extinction, $A_{K_s}$ as a
free parameter. In order to account for photometric uncertainties,
uncertainties of the line-of-sight distance, uncertainties in the
differential extinction correction of our photometry, and photometric
uncertainties intrinsic to the stars (e.g. due to variability of
differences in metal content), we include a smoothing parameter in the
fits, which is used to smooth the theoretical models with a
  Gaussian function of FWHM given by this parameter. It is let to vary
  freely, and is typically of the order one to a few.
 
\subsection{Number of star formation events}
\label{sec:number_events}

An important -- and a priori unknown -- ingredient in a stellar
population model fit to the KLF is the number of the star
formation events. In this work we are mainly interested in the
question whether stars are old enough to be dynamically relaxed, which
implies ages of at least a few Gyr. Since we are not interested in a high  precision of the  ages of given
events (which is very difficult to obtain), we
selected the ages that form part of our models after visual
inspection of the differences between single age KLFs \citep[see also][]{Nogueras-Lara:2019qd}. The older the
stellar population is, the slower the LF changes with age. We chose the
following 16 ages: 13, 11, 9, 7, 6, 5, 4, 3, 2, and 1  Gyr  and 800, 500,
250, 150, 100, 80, and 30 Myr. We chose Johnson-Cousins filters
  for the models. The differences between a Johnson-Cousins $K$ filter
and the $K_{s}$ filter used for the observations can be
neglected for the aims of this work.

\begin{figure}[!htb]
\includegraphics[width=\columnwidth]{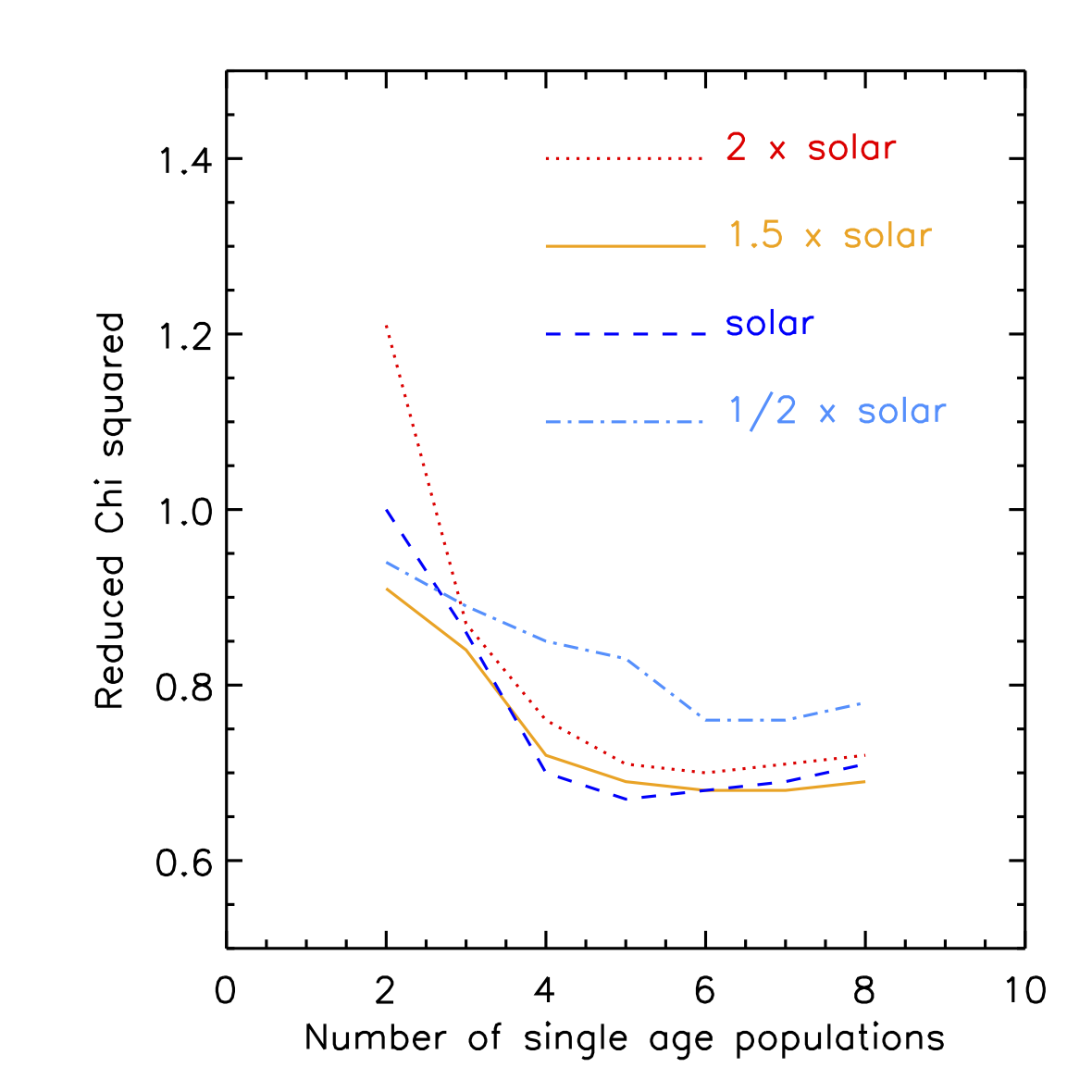}
\caption{\label{Fig:Chired_npop} Minimum reduced $\chi^2$ plotted
  against number of single age stellar populations that are used to
  fit the measured KLF. }
\end{figure}

\begin{figure}[!htb]
\includegraphics[width=\columnwidth]{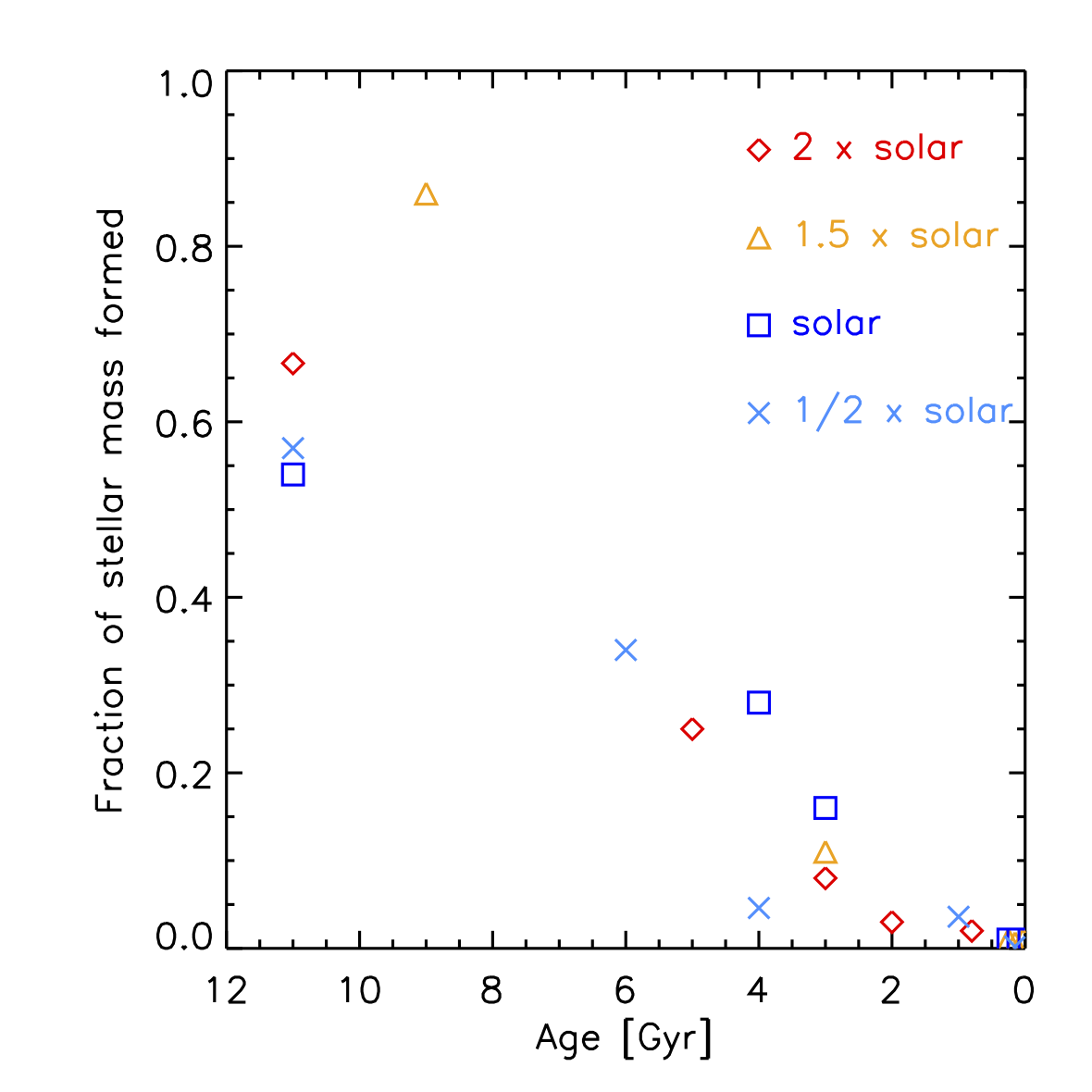}
\caption{\label{Fig:SFH_min} Star formation histories for five single
  age events with different mean metallicities. Not all data points
  can be clearly seen because they bunch together at young ages.}
\end{figure}

In a first step we estimated the number of single age populations
necessary to provide a good fit to the measured KLF. We fitted all
possible combinations of between two to eight theoretical LFs from our
initial age selection to the data and plotted the best reduced
$\chi^2$ against the number of populations, as shown in
Fig.\,\ref{Fig:Chired_npop}.  All fits with solar and supersolar
  metallicities reach very similarly low levels of reduced $\chi^2$
  with just five populations of different ages. Reduced $\chi^2$
  values for half solar metallicity are consistently higher than for
  the higher metallicities.  The best fit solutions was searched for
with the IDL MPFIT package \citep{Markwardt:2009fk}.  We set the
initial weight of all populations to zero except for the two oldest
populations that were assigned ten different initial random values,
with the best solution given by the lowest achieved $\chi^2$. We
assured through extensive tests (using thousands of random initial
values, assigning random or constant initial weights) that with this
setup the best-fit solutions were robust and did not correspond to
local minima.

\begin{figure}[!htb]
\includegraphics[width=\columnwidth]{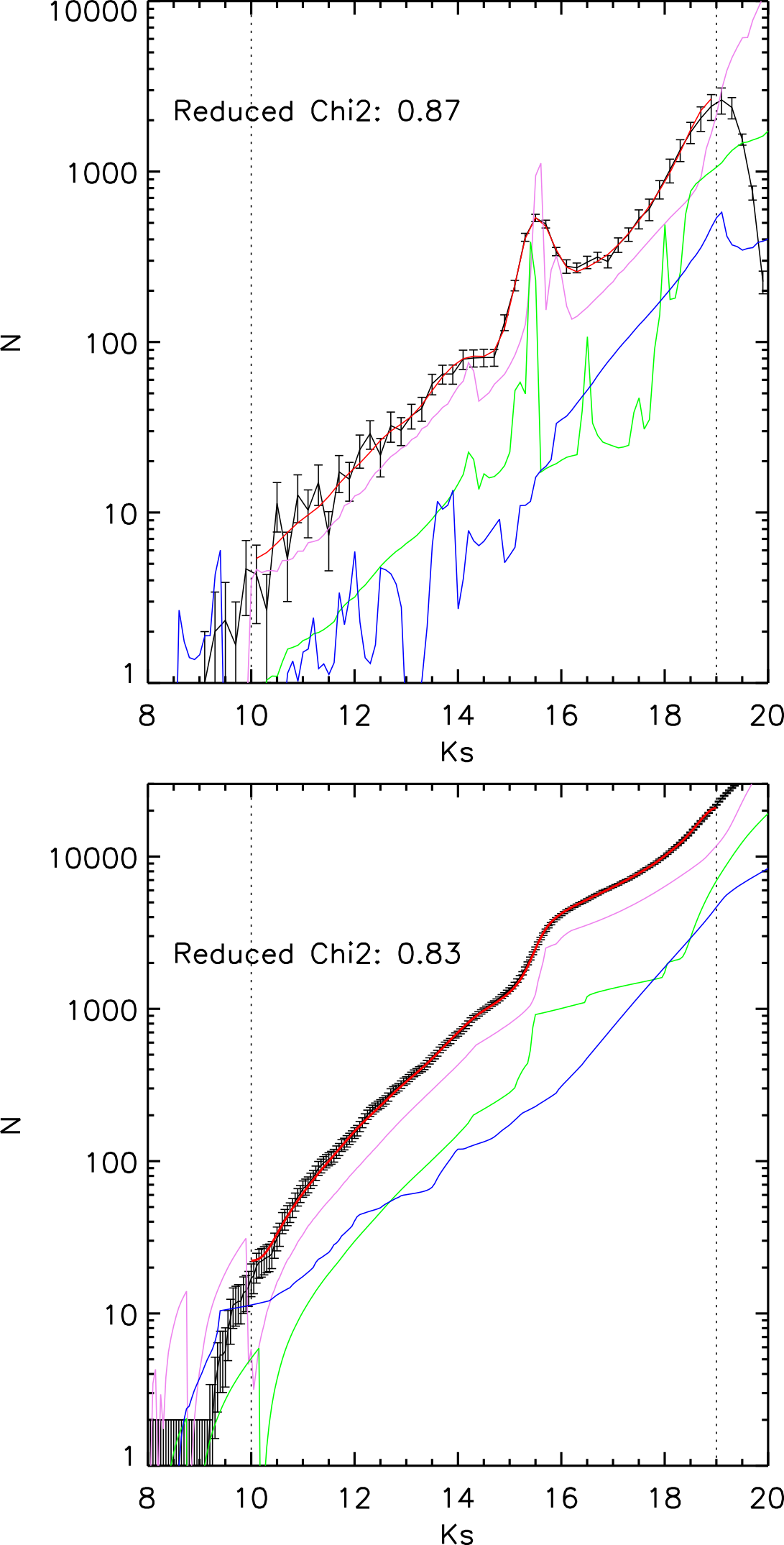}
\caption{\label{Fig:SFH_BEST} Top: Best fit of the KLF (foreground
    sources excluded, corrected for differential extinction and for
    completeness due to crowding) with 16 single age
  populations for $1.5$ times solar $[Fe/H]$ (column four in
  Table\,\ref{Tab:best-fit}). Violet: 13, 11, 9, 7\,Gyr; green: 2 and
  3\,Gyr; blue: $\lesssim0.5$\,Gyr. Bottom: The same, but for the
  cumulative KLF ( column four in
  Table\,\ref{Tab:best-cumlf})).}
\end{figure}

\begin{table}
  \caption{Best model fits to the KLF, considering stellar populations
    of 16 different ages and different metallicities. The columns
    after line 4, list the fraction of the total initially formed
    stellar mass corresponding to each age.}
 \label{Tab:best-fit}
\centering
\begin{tabular}{llllll}
  \hline 
  Age (Gyr) &  $0.5\,Z_{\odot}$ &  $1\,Z_{\odot}$  &  $1.5\,Z_{\odot}$  &  $2\,Z_{\odot}$ & Mixed\\
  \hline
$\chi^{2}_{\mathrm red}$ & $1.12$ & $0.87$ & $0.82$ & $0.92$  & $0.92$ \\ 
$<A_{K_{s}}>\tablefootmark{a}$    & $2.54$ & $2.58$ & $2.44$ & $2.43$ & $2.44$ \\
FWHM$_{smooth}\tablefootmark{b}$    & $1.08$ & $1.26$ & $1.19$   & $1.16$  & $1.15$ \\
\hline
13     &    $0.129$         & $0.378$ & $0.267$  & $0.497$   &  \\
11     &    $0.450$         & $0.097$ & $0.456$  & $0.050$   & $0.72$ \\
 9      &    $0.000$         & $0.000$ & $0.000$  & $0.311$  & $0$ \\
 7      &    $0.025$         & $0.000$ & $0.134 $ & $0.033$  & $0.15$ \\
 6      &    $0.287$         & $0.000$ & $0.000$  & $0.000$  & $0$ \\
 5      &    $0.000$         & $0.000$ & $0.000$  & $0.000$  & $0$ \\
 4      &    $0.068$         & $0.345$ & $0.000$  & $0.000$  & $0$ \\
 3      &    $0.000$         & $0.155$ & $0.066$  & $0.075$  & $0.09$ \\
 2      &    $0.000$         & $0.000$ & $0.048$  & $0.012$  & $0.02$ \\
 1      &    $0.029$         & $0.000$ & $0.000$  & $0.000$  & $0$ \\
$0.8$   & $0.000$         & $0.000$ & $0.000$  & $0.000$  & $0$ \\
$0.5$   & $0.000$         & $0.000$ & $0.000$  & $0.006$ & $0$ \\
$0.25$ & $0.002$         & $0.012$ & $0.013$  & $0.006$ & $0.008$ \\
$0.15$ & $0.002$         & $0.001$ & $0.006$ & $0.002$ & $0.003$ \\
$0.1$   & $0.003$         & $0.007$ & $0.000$ & $0.006$ & $0.003$ \\
$ 0.08$& $0.004$         & $0.000$ & $0.012$ & $0.00$        & $0$ \\
$0.03$ & $0.000$         & $0.004$ & $0.000$ & $0.002$ & $0.008$ \\
\hline
\end{tabular}
\tablefoot{
   \tablefoottext{a}{Mean extinction..}
  \tablefoottext{b}{FWHM of Gaussian smoothing parameter in units of bins.}
}
\end{table}

\begin{table*}
  \caption{Best model fits to the {\it cumulative} KLF, considering stellar populations
    of 16 different ages and different metallicities, with and without
    spectroscopically confirmed young stars. The columns
    after line 4, list the fraction of the total initially formed
    stellar mass corresponding to each age.}
 \label{Tab:best-cumlf}
\centering
\begin{tabular}{llllllllllll}
  \hline 
  Age (Gyr) &  $0.5\,Z_{\odot}$ &  $1\,Z_{\odot}$  &  $1.5\,Z_{\odot}$  &  $2\,Z_{\odot}$  &  $1\,Z_{\odot}\tablefootmark{c}$  &  $1.5\,Z_{\odot}\tablefootmark{c}$  &  $2\,Z_{\odot}\tablefootmark{c}$  &  $1.5\,Z_{\odot}\tablefootmark{c,d}$   &  $1.5\,Z_{\odot}\tablefootmark{c,e}$ &  $1.5\,Z_{\odot}\tablefootmark{c,f}$&  Mixed\tablefootmark{c,g}\\
  \hline
$\chi^{2}_{\mathrm red}$                              & $2.28$ & $0.77$ & $0.83$ & $0.88$ & $0.92$ & $0.70$ & $0.73$ & $0.73$ & $0.60$ & $0.74$ & $1.18$\\ 
$<A_{K_{s}}>\tablefootmark{a}$                 & $2.55$ & $2.45$ & $2.49$ & $2.50$ & $2.50$ &  $2.52$ & $2.55$ & $2.50$ & $2.52$ & $2.51$ & $2.53$ \\
FWHM$_{smooth}\tablefootmark{b}$           & $3.48$ & $3.18$ & $3.48$  & $3.72$ & $2.84$ & $2.64$ & $2.91$ & $2.41$ & $2.36$ & $2.27$ & $4.09$ \\
\hline
13     &    $0.445$        & $0.823$ & $0.745$   & $0.680$   & $0.795$ & $0.821$ & $0.748$ & $0.833$ & $0.729$ & $0.805$ & $0.000$ \\
11     &    $0.000$        & $0.000$ & $0.000$   & $0.000$   & $0.000$ & $0.000$ & $0.000$ & $0.000$ & $0.000$ & $0.000$ & $0.000$ \\
 9      &    $0.000$          & $0.000$ & $0.000$   & $0.000$ & $0.000$ & $0.000$ & $0.000$ & $0.000$ & $0.000$ & $0.000$ & $0.652$ \\
 7      &    $0.000$          & $0.000$ & $0.105$   & $0.000$ & $0.000$ & $0.000$ & $0.000$ & $0.000$ & $0.000$ & $0.000$ & $0.000$ \\
 6      &    $0.368$          & $0.000$ & $0.000$   & $0.165$ & $0.010$ & $0.000$ & $0.084$ & $0.000$ & $0.000$ & $0.000$ & $0.116$ \\
 5      &    $0.101$          & $0.000$ & $0.000$   & $0.000$ & $0.000$ & $0.000$ & $0.000$ & $0.000$ & $0.000$ & $0.000$ & $0.000$ \\
 4      &    $0.000$          & $0.000$ & $0.000$   & $0.023$ & $0.000$ & $0.000$ & $0.000$ & $0.000$ & $0.000$ & $0.086$ & $0.000$ \\
 3      &    $0.000$          & $0.15$ & $0.124$     & $0.102$ & $0.144$ & $0.131$ & $0.116$ & $0.127$ & $0.187$ & $0.067$ & $0.129$ \\
 2      &    $0.050$          & $0.000$ & $0.000$   & $0.000$ & $0.000$ & $0.000$ & $0.000$ & $0.000$ & $0.000$ & $0.000$ & $0.000$ \\
 1      &    $0.000$          & $0.000$ & $0.000$   & $0.000$ & $0.013$ & $0.002$ & $0.002$ & $0.014$ & $0.013$ & $0.002$ & $0.000$ \\
$0.8$   & $0.000$          & $0.000$ & $0.000$   & $0.000$ & $0.000$ & $0.005$ & $0.003$ & $0.000$ & $0.000$ & $0.000$ & $0.000$ \\
$0.5$   & $0.009$          & $0.004$ & $0.004$   & $0.006$ & $0.003$ & $0.035$ & $0.039$ & $0.020$ & $0.036$ & $0.035$ & $0.005$ \\
$0.25$ & $0.008$          & $0.009$ & $0.008$   & $0.009$ & $0.003$ & $0.002$ & $0.001$ & $0.003$ & $0.004$ & $0.001$ & $0.015$\\
$0.15$ & $0.003$          & $0.001$ & $0.002$   & $0.003$ & $0.000$ & $0.000$ & $0.000$ & $0.000$ & $0.000$ & $0.000$ & $0.003$ \\
$0.1$   & $0.000$          & $0.007$ & $0.004$   & $0.006$ & $0.000$ & $0.000$ & $0.002$ & $0.000$ & $0.000$ & $0.000$ & $0.000$ \\
$ 0.08$& $0.015$          & $0.000$ & $0.003$   & $0.003$ & $0.006$ & $0.003$ & $0.002$ & $0.002$ & $0.005$ & $0.002$ & $0.013$ \\
$0.03$ & $0.001$          & $0.005$ & $0.004$   & $0.004$ & $0.001$ & $0.002$ & $0.002$ & $0.001$ & $0.002$ & $0.002$ & $0.004$ \\
\hline
\end{tabular}
\tablefoot{
   \tablefoottext{a}{Mean extinction..}
  \tablefoottext{b}{FWHM of Gaussian smoothing parameter in units of bins.}
  \tablefoottext{c}{Spectroscopically identified young stars excluded.}
  \tablefoottext{d}{Assumption of top heavy IMF.}
  \tablefoottext{e}{Fit limited to magnitudes brighter than $K_{s}=18.75$.}
  \tablefoottext{f}{Fit limited to magnitudes brighter than $K_{s}=19.25$.}
  \tablefoottext{g}{Assumption of mean metallicity $[Fe/H]=0.5$ solar
    for stars older than 10\,Gyr, $[Fe/H]=1.5$ solar for younger stars.}
}
\end{table*}

In Fig.\,\ref{Fig:SFH_min} we show the SFH for assuming five single
age star forming events and for metallicities of $0.5$, $1$,
  $1.5$, and $2$ times solar. In all cases, $>50\%$ of the original
  stellar mass formed $\gtrsim10\,$Gyr ago. In the solar metallicity
  model about 40\% of the stellar mass may have formed at intermediate
  (3-4 Gyr) ages. Spectroscopic studies indicate that about 80\% of
  the stellar mass formed more than 5\,Gyr ago
  \citep{Blum:2003fk,Pfuhl:2011uq} and that the mean metallicity in
  the NSC is super-solar
  \citep{Do:2015ve,Schultheis:2019lw}. Therefore, we prefer the models
  with $1.5$ to $2$ times solar metallicity.

\subsection{Fits with many populations}

Next, we determined the best fit solution by allowing a
  combination made up of all 16 single age populations for which we
  had calculated the KLF.  We set the initial weight of all
  populations to zero except for the five oldest populations that were
  assigned one thousand different initial random values. We ensured by
  extensive tests that we could find the global $\chi^{2}$ minimum
  successfully with this approach. Table\,\ref{Tab:best-fit} gives the
  values of the parameters for the best fits for the models with mean
  metallicities of $0.5$, $1.0$, $1.5$, and $2$ times solar as well as
  for a model with a mixture of metallicities for the three oldest
  ages (20\% for $0.5$, 20\% for $1$, 30\% for $1.5$, and 30\% for $2$
  times solar). The resulting best-fit parameters for the latter are
  close to the case of $1.5$\,times solar metallicity, as
  expected. This test served to verify the validity of our approach to fit
  mean metallicity models to a stellar population that contains in
  reality a range of different
  metallicities. The top panel of figure\,\ref{Fig:SFH_BEST} shows the best fit for the
  $1.5$ solar metallicity case, that results in the overall lowest
  $\chi^{2} $.  The
  formal uncertainties of the parameters are not listed in the
  table. They are generally large because of the degeneracy of the
  high similarity of KLFs between adjacent age bins.

 While binned data have the advantage of being easier to
  interpret, the binning may impose a loss of information or
  additional uncertainty. We therefore also performed fits on
  cumulative KLFs.  The resulting best-fit parameters are listed in
  Table\,\ref{Tab:best-cumlf}. The bottom panel of
  Fig.\,\ref{Fig:SFH_BEST} shows the fit of the cumulative KLF for
  $[Fe/H] = 1.5$ solar. The fits to the cumulative LF are more stable
  to initial conditions and are more consistent in the derived SFH for
  different metallicities.  Therefore, in the following part of this
  work, we focus on the results from fits to the cumulative KLF
  from this point on.

%  {\bf For super solar $[Fe/H]$, a clear picture emerges: At least
%    80\% of the stellar mass formed $7-11$\,Gyr ago; no detectable
%    star formation occurred at intermediae ages of $4-6$\,Gyr ago;
%    $\sim$10\% of stellar mass formed at intermediate ages $2-4$\,Gyr
%    ago; no detectable star formation occurred $1\,$Gyr ago; finally a
%    few percent of the stellar mass fored $100-500$\,Myr and
%    $<100$\,Myr ago.}

\subsection{Systematics}

There are many potential sources of systematic uncertainty in the
stellar population fits. In this section we examine the ones that we
could identify and considered possibly significant. By using the
cumulative LFs we can already eliminate binning as a possible source
of systematics. Other sources are: 

\begin{enumerate}

\item {\it Foreground stars, IRS\,13}: Systematic effects may result from our
choice of threshold for exclusion of foreground stars and from the
inclusion or exclusion of the potential cluster IRS\,13
\citep[e.g.][]{Schodel:2005le,Fritz:2010tk}, We created the KLF of the
large field with a stricter criterion for the exclusion of foreground
stars ($H-K{s}<1.7$). The resulting KLF is indistinguishable, within
the errors, from the one used in the analysis above. The inclusion or
exclusion of sources from the IRS\,13 region into the KLF did not have
any significant effect.

\item {\it Spectroscopically identified massive, young stars}.  It is
well established that there is a few million year old population of
massive young stars present within a projected distance of $0.5$\,pc
from Sgr\,A*. The properties and formation of these stars are still a
topic of intense investigation. There is now strong evidence that the
corresponding starburst must have had a top-heavy initial mass
function
\citep{Nayakshin:2005ve,Bartko:2010fk,Pfuhl:2011uq,Lu:2013fk}. This
means that the contamination of the KLF by lower-mass young stars from
this event, which have not been spectroscopically identified, is
probably minor \citep[see Figure\,12 in][]{Gallego-Cano:2018nx}. 
  Evidence for this latter point is also provided by our comparison of
  the KLFs inside and outside a projected radius of $R=0.5$\,pc from
  Sgr\,A*, as shown in Fig.\,\ref{Fig:KLF_large}. In our modelling we
  have not included any population of stars in the corresponding age
  range, in particular because BaSTI models do not extend to such
  young ages. We therefore cleaned the LFs from the spectroscopically
  identified young, massive stars. The resulting best-fit parameters
  are listed in the columns after column 5 in
  Tab.\,\ref{Tab:best-cumlf}. The reduced $\chi^{2}$ values become
  somewhat smaller and super-solar metallicities become somewhat  more
  favoured. The general SFH is not changed. We conclude that the
  presence of young, massive stars does not introduce any strong bias,
  but that it is better to remove those stars from the sample. From
  here on, we therefore only use LFs without the spectroscopically
  identified young, massive stars. 

\item {\it Initial mass function}: We used a Salpeter
  IMF, that is, an exponential IMF with an exponent of $-2.35$.  Our
data are dominated by giants which cover a very narrow mass range,
(e.g.\ $0.7-1.7$\,M$_{\odot}$ for 2\,Gyr age, $0.7-1.1$\,M$_{\odot}$ for
9\,Gyr age). Therefore there is little leverage to measure the
IMF. Also, the SFH study by \citet{Pfuhl:2011uq} finds that a
Chabrier/Kroupa IMF (very similar to a Salpeter IMF at masses above
one solar mass) works well on the central parsec, except for the
already discussed case of the young, massive stars. We tested the
  assumptions on the IMF by fitting the cumulative KLF with BaSTI
  isochrones with a top-heavy IMF with an exponent of $-1.7$. The
  resulting SFH is detailed in column nine of
  Table\,\ref{Tab:best-cumlf}. Within the uncertainties it is
  indistinguishable from the one derived assuming a Salpeter IMF.
\item {\it Fitting range}: In order to test the importance of the faint
magnitude cut-off used to fit the cumulative KLF -- for example
  because of potential completeness variation due to spatially
  variable extinction --, we performed fits with cut-offs at
  $K_{s}=18.75$ and at $K_{s}=19.25$ instead of at $K_{s}=19$. 
    The resulting SFHs are listed in columns ten and eleven of of
    Table\,\ref{Tab:best-cumlf}. The brighter cut-off results in a
    somewhat stronger contribution appearing at 3\,Gyr, while the
    fainter cut-off will increase the weight of the oldest population
    and shift the age of the 3\,Gyr population a bit towards
    4\,Gyr. The changes are not substantial and reflect the fact that
  the oldest population are traced best by the increase of star counts
at the faintest magnitudes.

\item {\it Metallicity}: Most spectroscopic studies in the past decades
found solar to supersolar mean metallicities at the GC
\citep{Ramirez:2000ys,Cunha:2007oq,Do:2015ve,Feldmeier-Krause:2017kq,Nandakumar:2018zr}.
Some studies found somewhat lower mean metallicities
\citep{Rich:2017rm}. The meta-study by \citet{Schultheis:2019lw}
argues that the metallicity distribution in the innermost bulge is
complex, but that the GC has a high supersolar mean metallicity with
$[Fe/H]=+0.2$\,dex.  We note that the latter study uses GC sample
  of \citet{Nandakumar:2018zr}, who did not find any stars with
  sub-solar metallicity in the proper GC field.  Finally, the
photometric study by \citet{Nogueras-Lara:2018oe} of the innermost
bulge, at projected distances of $60-90$\,pc to Galactic north from
Sgr\,A*, indicates a single-age stellar population of twice solar
metallicity. In conclusion, current evidence points towards solar
  to  super-solar mean metallicity in the nuclear star
  cluster.

  The stellar population fits to the KLFs performed here, are
  consistent with the spectroscopic results, as can be seen in
  Fig.\,\ref{Fig:Chired_npop} and Tab.\,\ref{Tab:best-cumlf}. We note
  that exclusion of the spectroscopically identified massive stars
  improves the $\chi^{2}$ of the fit with supersolar metallicities,
  but not for the one with solar mean metallicity. A mean metallicity
  of half solar for all ages can be excluded with high confidence and
  the overall best fit is achieved for a metallicity of
  $[Fe/H]=+0.2$\,dex, after exclusion of the spectroscopically
  identified young stars. Finally, we have performed a population fit
  in which we assign half solar metallicity only to the oldest two age
  bins (13 and 11\,Gyr). The result is shown in the last column of
  Tab.\,\ref{Tab:best-cumlf}: Reduced $\chi^{2}$ is higher than for
  all other cases except the one when all populations are forced to
  have low metallicity. In fact, the weights of the low metallicity
  populations result to be zero, which is equivalent to limiting the
  age of the stars in the cluster. At the same time the best-fit
  smoothing factor is higher than in all other cases. We conclude that
  setting the oldest populations to a low metallicity does not lead to
  a satisfactory fit.  To conclude, as concerns metallicity, the
  inferred star formation
  historiesis robust with respect to the assumed mean metallicity in the
  range one to two times solar $[Fe/H]$.
\item {\it Stellar evolutionary codes:} We also fitted the data with
  theoretical cumulative KLFs based on two other stellar evolution
  codes: PARSEC \citep[release v$1.2$S $+$ COLIBRI
  S$\_35$][]{Bressan:2012xy,Chen:2014nr,Chen:2015sf,Tang:2014rm,Marigo:2017fb,Pastorelli:2019er}
  and MIST
  \citep{Dotter:2016sh,Choi:2016qh,Paxton:2011vh,Paxton:2013vb,Paxton:2015hz}. We
  used scaled solar metallicities.  The parameters of the best fits are
  listed in Tables\,\ref{Tab:PARSEC} and \ref{Tab:MIST} in
  appendix\,\ref{sec:models}. We note:

\begin{itemize}
\item The  achieved minimum $\chi^{2}$ values are worse for PARSEC and MIST
  than for BaSTI. More importantly, while for BaSTI the minimum
  $\chi^2$ values decrease further when we exclude the
  spectroscopically young, massive stars from the data, as is expected
  because our models do not contain sufficiently young isochrones, the
  $\chi^{2}$ values actually {\it increase} in this case for MIST and
  PARSEC to about twice the minimum values reached in BaSTI. It
  appears that the most recent release of BaSTI \citep{Hidalgo:2018mw}
  is suited better for the GC stellar populations than the other two models.
\item Independent of metallicity and whether the young massive stars
    are excluded or included, all SFHs derived with PARSEC imply that
    $\gtrsim40\%$ of stars formed 5\,Gyr or less ago. This result contradicts
    constraints from spectroscopic studies, which require the stellar mass
    formed at these ages to be $\lesssim20\%$
    \citep{Blum:2003fk,Pfuhl:2011uq}.
  \item The SFH inferred with MIST isochrones after exclusion of
    spectroscopically identified massive, young stars agrees for solar
    and $1.5$ solar $[Fe/H]$ -- in spite of the high $\chi^2$ values
    -- reasonably well with the SFH from BaSTI isochrones at these
    metallicities. The main difference is that the stellar mass
    formed at intermediate ages is concentrated at 3\,Gyr in the BaSTI
    fits and split between 4 and 2\,Gyr in the MIST fits. Some age
    shifts between the models are expected \citep[see also the
    comparison of MIST and BaSTI LFs discussed in the appendix
    of][]{Nogueras-Lara:2019qd}.
\end{itemize}

\end{enumerate}

{\it Conclusion}: The systematic effects studied
  here show that the mean metallicity of the NSC appears to be solar
  to super solar, with only small differences in the derived SFH. In
  agreement with spectroscopic studies, sub-solar mean metallicity
  appears to be highly unlikely. After excluding fits with PARSEC
  isochrones on the basis that they do not agree with the mentioned
  spectroscopic findings and do not agree with the fits based on MIST
  or BaSTI isochrones, we conclude that none of the examined
  systematics alters the derived SFH significantly.

\begin{figure}[!htb]
\includegraphics[width=\columnwidth]{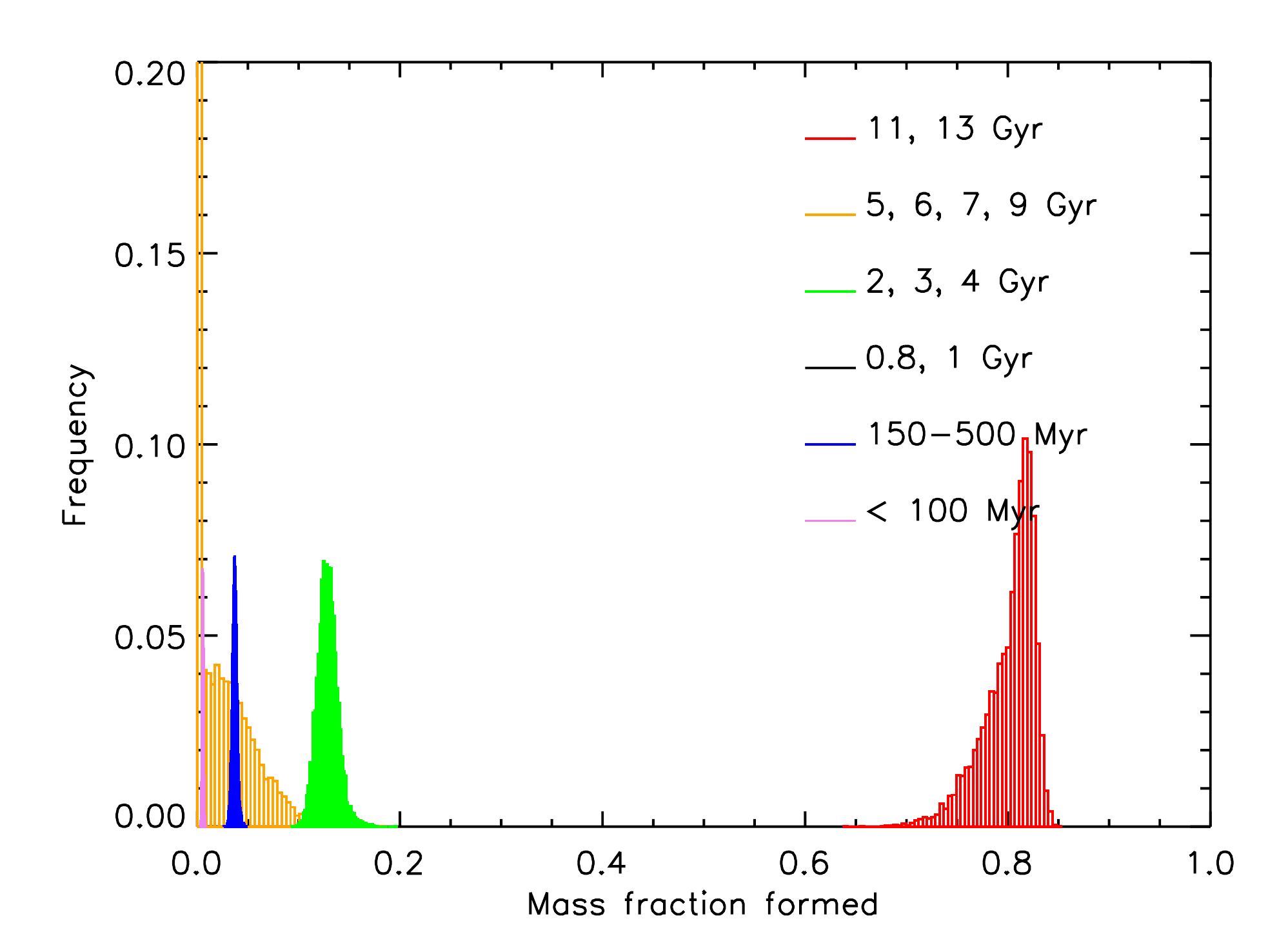}
\caption{\label{Fig:SFH_MC} Histograms of the stellar mass fraction
  formed in given age ranges derived from fits to MC simulations of
  the cumulative KLF with BaSTI isochrones and assuming a mean $[Fe/H]$
  of $1.5$ times solar.}
\end{figure}

\subsection{Monte Carlo simulation} 
\label{sec:MC}

To further explore uncertainties, we created $10,000$
realisations of the cumulative KLF in a Monte Carlo (MC) simulation,
based on the mean and standard deviation in each magnitude bin, and
fitted those random cumulative KLFs again with a combination of 16
different populations. Spectroscopically confirmed massive, young
stars were excluded.  The fits were performed for mean $[Fe/H]$
  of $1.0$, $2.0$, and $1.5$ times solar.  The lowest reduced $\chi^2$
  values were achieved for $[Fe/H]=2.0$\,solar in $26$\% of the cases,
  for $[Fe/H]=1.5$\,solar in $74$\% of the cases, and for
  $[Fe/H]=1.0$\,solar in $0$\% of the cases, indicating a preference
  for super solar mean metallicity of roughly $+0.2$\,dex, in
  agreement with the spectroscopic analysis by
  \citet{Schultheis:2019lw}. Figure\,\ref{Fig:SFH_MC} show
histograms of the relative contributions of the different ages for a
mean metallicity of $1.5$ times solar $[Fe/H]$. We show the
corresponding plot for BaSTI isochrones and solar metallicity and for
MIST isochrones and $1.5$ times solar metallicity in
appendix\,\ref{app:MC}, too. They show broader, less well defined
distributions, but agree well with Fig.\,\ref{Fig:SFH_MC} in
general. In appendix\,\ref{app:MC} we also show histograms of the
reduced $\chi^{2}$ values for the MC fits with BaSTI isochrones and
different mean metallicities.

\subsection{Conclusions on the star formation history}

{\it Star formation history}: After having examined systematics
  and having performed the MC simulations, a clear picture of the
  formation history of the MWNSC emerges:
\begin{enumerate}
 \item About 80\% or of the
  star formation occurred $>10$\,Gyr ago.
\item Very little activity
occurred between 5 and 10\,Gyr ago, where at most 5\%, but probably
close to 0\% of star formation occurred. 
\item There is clear evidence for
an intermediate age population of $2-4$\,Gyr, which {\bf contributes
  around 15\%} of star formation. 
\item A clear minimum existed at around
1\,Gyr. 
\item Finally, a few percent of the nuclear cluster's mass formed
in the past 150-500\,Myr.
\item $\lesssim 1\%$ of star formation happened in the past 100\,Myr,
  with the caveat that our models do not include populations younger
  than 30\,Myr and that the spectroscopically confirmed stars
  pertaining to the youngest star formation event were excluded from
  the fits.
\end{enumerate}

\section{Stellar density near Sgr\,A*}

\begin{figure}[!htb]
\includegraphics[width=\columnwidth]{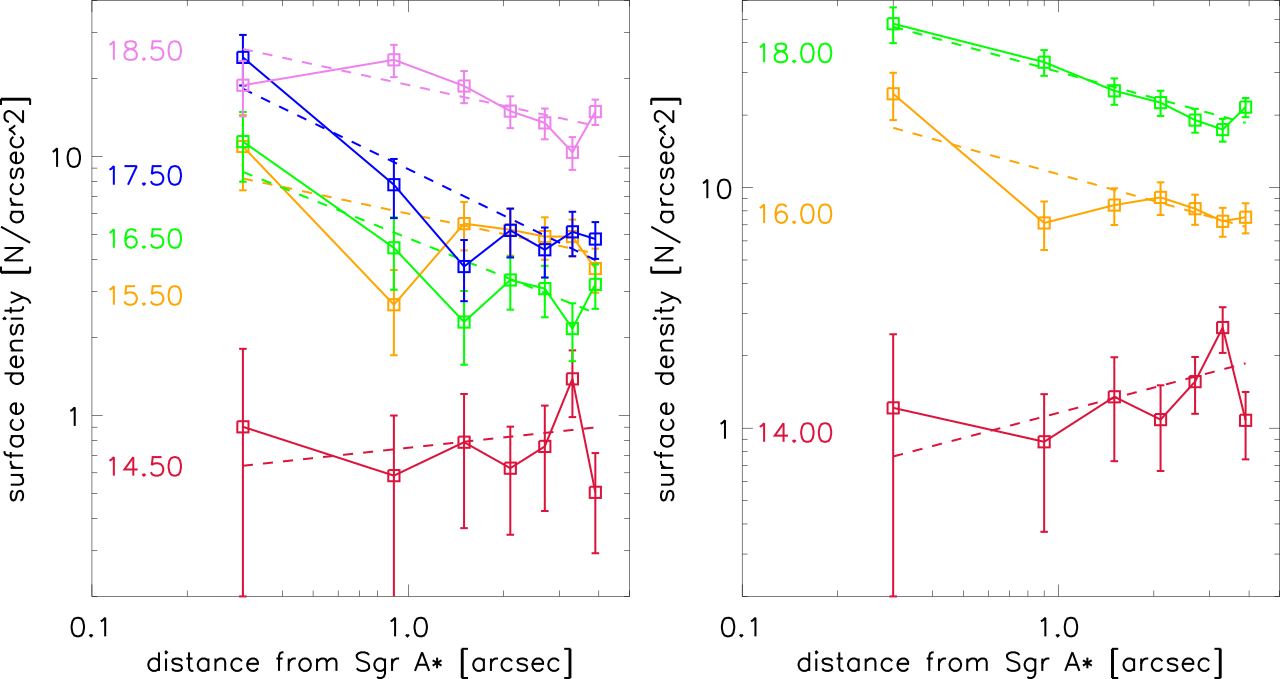}
\caption{\label{Fig:SFD} Projected stellar surface density  in the
  deep field. Left: One-magnitude wide bins. Right: Two-magnitude wide
  bins (the brightest bin contains all stars $K_{s}<15$. The dashed lines
  show simple power-law fits.}
\end{figure}

We used the high quality central field data to infer the profile of
the stellar surface number density within $5"$, or $0.2$\,pc, of
Sgr\,A*. The aim is to look for the signature of a relaxed stellar
cusp, which can only be traced by stars old enough to have undergone
dynamical relaxation. Since stars at different mean magnitudes can
have different mean ages, we plot the analysis for different magnitude
bins (Fig.\,\ref{Fig:SFD}). Extinction and completeness correction
were applied and spectroscopically identified young hot stars were
excluded. Power laws were fitted to the data points. The different
power-law indices are listed in Table\,\ref{Tab:sfd_alphas}.

All stars in magnitude bins fainter than $K_{s}=15$ show a density
increase towards Sgr\,A* that can be approximated by a power-law.  The
power-law indices are consistent with each other within the
uncertainties. Stars in the brightest bin do not show this increase,
but are rather consistent with a constant, or even decreasing, surface
density. The stellar surface densities presented here reach about
0.5\,mag deeper than the ones presented in
\citet{Gallego-Cano:2018nx}.

\begin{table}
\caption{Power-law index $\alpha$  of a single power-law fitted to the projected
surface density of stars around Sgr\,A* in one-magnitude wide
brightness bins. \label{Tab:sfd_alphas}
}
\begin{center}
\begin{tabular}{ll}
\hline
\hline
$K_{s}$  &  $\alpha$\\
\hline
14.5 & $0.13\pm0.32$\\
15.5 & $-0.26 \pm- 0.15$\\
16.5 & $-0.49\pm 0.14$\\
17.5 & $-0.59\pm0.14$\\
18.5 & $-0.27\pm0.11$\\
\hline
\end{tabular}
\end{center}
\end{table}

\section{Discussion}

\subsection{Stellar population and formation history}

The photometric measurements for $>30,000$ stars, combined with
extinction measurements with an angular resolution
around $1"$, allow us to present the so far deepest and most
detailed KLF of the Milky Way's NSC. 

An important fundamental question is the approximate distance depth of
the data, that determines which stellar structure dominates our
sample. As concerns stars in the Galactic foreground, located in
spiral arms or in the bar/bulge in front of the central molecular
zone, they can be reliably excluded by their blue colours as we do in
this work and as can be seen in the detailed colour magnitude diagrams
presented in previous work \citep[see , e.g.\ ,
][]{Schodel:2010fk,Nogueras-Lara:2019yj}. It is less straightforward
to assess the contribution from background stars. However, a look at
the surface density of stars in the GC region can help. The surface
flux and number density as a projected function from Sgr\,A* show that
the projected stellar density within a projected radius
$R\lesssim1\,$pc is at least five times higher than at $R\gtrsim5\,$pc
because of the steep increase of stellar density towards the cusp of
the Milky Way's nuclear star cluster \citep[e.g.\
][]{Schodel:2014fk,Gallego-Cano:2018nx}. Therefore, we assume here
that our KLF is dominated by stars inside the central few parsecs of
the MWNSC, with a contamination of not more than about 10\% of stars
in the fore- and background of the cluster. Also, our observation of
an indistinguishable KLF between the very central parts of the MWNSC
($R<0.5$\,pc) and the parts at somewhat greater $R$, as shown in
section\,\ref{sec:KLF_central} supports our assumption that our sample
is dominated by stars within a few parsecs around Sgr\,A*.

The KLF appears to be well mixed and shows hardly any kind of
dependence on distance from Sgr\,A*. There is some excess in the
numbers of stars brighter than $K_{s}\approx14.5$ at projected
distances $R<12.5"$/$0.5$\,pc from Sgr\,A* due to the presence of
young, massive stars , but this excess is small. At magnitudes
$K_{s}>15$ the KLF in the centralmost arcseconds is identical, within
the uncertainties, to the one at distances of $R=25"$/1\,pc. The
stellar population therefore appears to be homogeneous and we can
analyse the SFH history by using the KLF of the entire field.

Because of the sensitivity of our data we detect an upturn of the KLF
at $K_{s}>18$. While previous, shallower KLFs, as presented in
\citet{Genzel:2003it}, \citet{Schodel:2007tw}, or
\citet{Pfuhl:2011uq}, were consistent with an old, bulge-like stellar
population with an admixture of young stars, the KLF presented here
requires the presence of intermediate-age stars of ages 2-4\,Gyr.

This is the first time that a detailed SFH of the MWNSC has been
inferred from its KLF. Although the KLF of the MWNSC is included in
the data of \citet{Figer:2004fk} it is not discussed {\bf separately}
there. Other publications, e.g.\ \citet{Genzel:2003it} and
\citet{Schodel:2010fk} present KLFs for the central parsec of the GC,
but the KLFs are shallower and less detailed. Also, they do not
discuss the SFH apart from noting that the KLF is consistent with an
old, bulge-like population. The deep, carefully extinction-corrected
KLF presented here contains features that are sensitive to the
SFH. Those are: (1) The number of stars brighter than the AGB are
sensitive to recently formed massive stars, (2) the relative weights
and distances between the RC and the RGBB (Red Giant Branch Bump) are
sensitive to age and metallicity of stars older than a few Gyr (RC and
RGBB are not individually resolved in our KLF but are mixed together
in the RC bump), (3) the upturn of the KLF at magnitudes fainter than
the RGBB (intermediate age stars), and (4) the number counts in the
region between the upturn of the KLF and the RC/RGBB bumps (young
stars). {\it Finally through the use of the cumulative KLF we can minimise
  potential systematic effects due to the choice of the bin size in
  the differential KLF. }

 {\it After investigating the possible systematics, the following
  general picture emerges:} (1) About $80\pm5\%$ of originally formed
stellar mass formed very early in the Milky Way's history. In all
cases the model LFs fitted to the data prefer ages of 13\,Gyr. We are
certain that such an old age must be interpreted with certain caution
given that stellar evolutionary models may give rise to ambiguities at
such old ages and high metallicities, where they cannot be calibrated
well. We therefore believe that it may be safe to say that 
$\sim$80\% of the original cluster mass formed at least about 10\,Gyr
ago. (2) Very little star formation occurred at
intermediate-old ages of 5 - 9\,Gyr, at most a few percent, with zero
being the most likely scenario. (3) Star formation clearly occurred at
intermediate-young ages of 2-4\,Gyr, when roughly 15\% of the
originally created stellar mass formed. The age uncertainty is largely
model-dependent. If we rely on any given evolutionary track then this
star formation event appears to be clearly placed in time, for example
$3.0\pm0.5$\,Gyr when relying on BaSTI isochrones. (4) Star formation
activity was close to zero about 1\,Gyr ago (see
Table\,\ref{Tab:best-fit}). (5) Finally, there is evidence of star
formation in the past few 100\,Myr and in the recent past ($<100$\,Myr
ago), although we cannot pinpoint the exact ages. About 1\% of the
stars formed in the past 100\,Myr, and of the order 3\% between 150
and 500 Myr ago. Our elaborate multi-population fitting therefore
confirms that the star formation history of the NSC can be fitted well
with a small number of star formation events: One at $\gtrsim10\,$Gyr,
one at $\sim$3\,Gyr and about 3 at 0-500\,Myr, as shown in section\,\ref{sec:number_events} (see
Figure\,\ref{Fig:Chired_npop}). More star formation events are needed
at young ages because of the rapid evolution of the KLF of at ages
younger than about 1\,Gyr.

The SFH derived here agrees well with the most recent spectroscopic
determination of the SFH by \citet{Pfuhl:2011uq} or with previous work
by \citet{Blum:2003fk}. Those publications found that of the order
70-80\% of the stars in the NSC formed more than 5\,Gyr ago.  We
  note, however,  that: (a) Our work on the KLF has
pushed back the age of the NSC considerably because of the sensitivity
of the KLF to the onset of the main sequence of old stellar
populations. (b) It appears that there was just a single star formation
period at intermediate age, around 3\,Gyr.  We can
also confirm the global star formation minimum that
\citet{Pfuhl:2011uq} found for ages $\sim$1\,Gyr. 

\begin{figure}[!htb]
\includegraphics[width=\columnwidth]{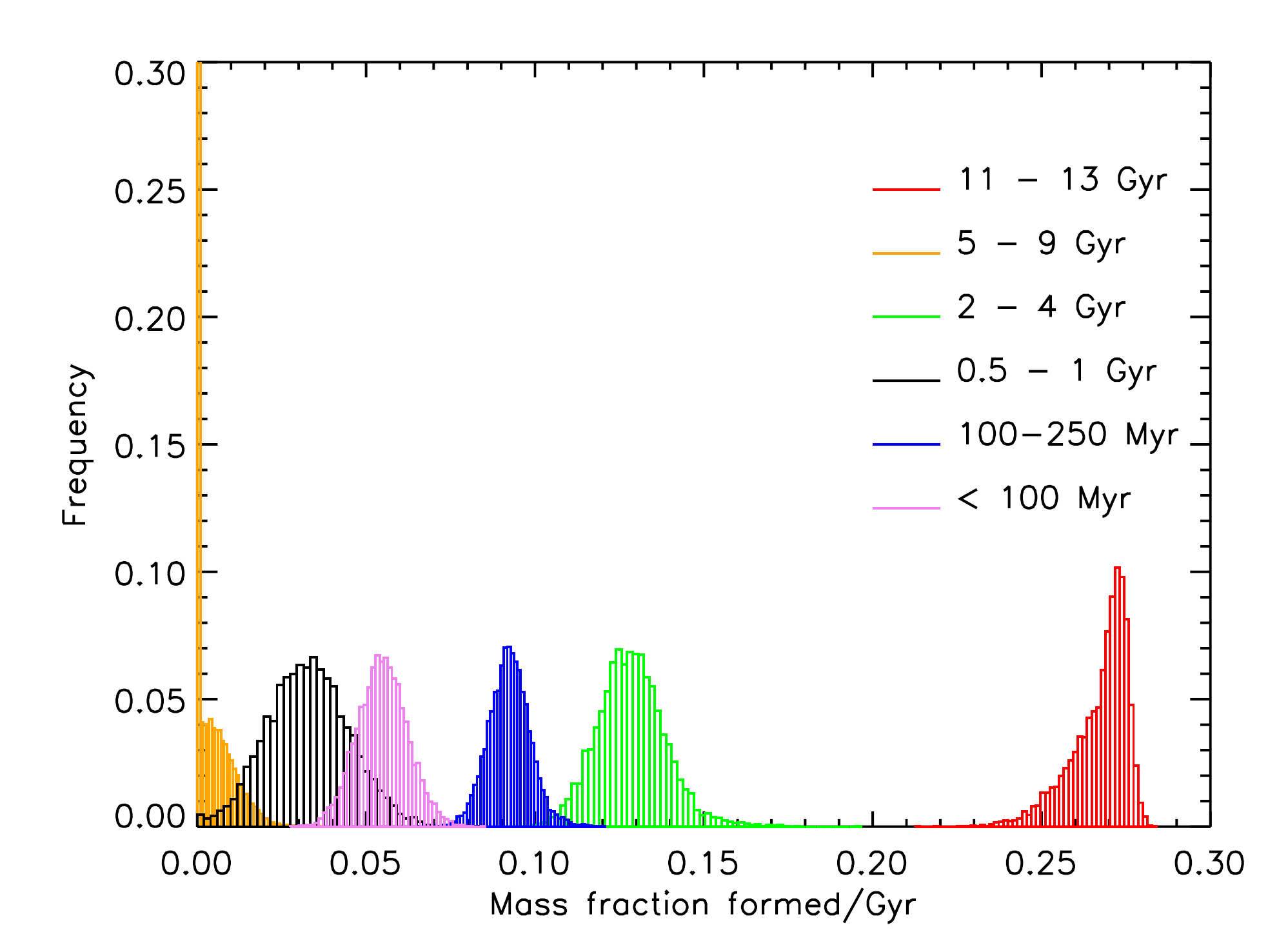}
\caption{\label{Fig:SFR} Relative star formation rate in the NSC as a
  function of time. Histograms corresponding to the SFH from the lower
  panel of Fig.\,\ref{Fig:SFH_MC} divided by the approximate length
  of the corresponding time intervals. }
\end{figure}

In Fig.\,\ref{Fig:SFR}, we present the star formation rate on a
  linear scale, by dividing the stellar mass fractions formed in a
  given time by the approximate length of the time interval. For the
  latter we assumed 3\,Gyr for the oldest event, 5\,Gyr for the
  quiescent time 5-9 Gyr ago, 1\,Gyr for the intermediate age stars
  and correspondingly shorter (and better defined) time intervals for
  younger events. In agreement with the SFR derived from spectroscopic
  work by \citet{Pfuhl:2011uq} we see that the star formation rate
  decreased until 1\,Gyr ago but then increased again. Contrary to the
  model presented in \citet{Pfuhl:2011uq} we find a long minimum of
  the SFR at ages roughly 5 - 9 Gyr ago. For modelling purposes the SFH
  of the NSC can be simply reproduced by assuming five different
  bursts of star formation.  Minima similar to the one observed at
$\sim$ 1\,Gyr may have occurred at other times because star formation
will probably not have been homogeneous in time and have proceeded in
bursts separated by tens or hundreds of millions of years. The one at
0.8-1\,Gyr can, however, be easily detected because the RC magnitude
shows a very clear minimum brightness at ages around 1\,Gyr that lasts
for at most a few 100\,Myrs \citep{Girardi:2016fk}. Combined with the
large number of stars in the RC, we would be very sensitive to any
star formation event that occurred about 1\,Gyr ago.

Our SFH fits prefer high metallicities. This agrees with the
most recent spectroscopic analyses \citep{Schultheis:2019lw}. Hence,
the assumption of supersolar metallicity of NSC appears to be correct.

\subsection{Stellar cusp around Sgr\,A*}

The presence of a density cusp around Sgr\,A* is expected for stars
older than the dynamical relaxation time. Relaxation by two-body
effects takes of the order 10\,Gyr in the NSC, but several effects,
such as the presence of massive perturbers or mass segregation can
speed this process up significantly
\citep{Perets:2007ud,Preto:2010kx,Alexander:2017fk}. It appears
therefore reasonable to assume that stars older than a few Gyr are
dynamically relaxed at the GC. The expected stellar cusp is not
observed for bright giants, but has been confirmed to be present for
stars fainter than $K_{s}\approx15.5$
\citep{Yusef-Zadeh:2012pd,Gallego-Cano:2018nx,Schodel:2018db,Habibi:2019tl}. Here
we present even deeper star counts than in \citet{Gallego-Cano:2018nx}
and confirm that the cusp is consistently detected in the form of a
power-law increase of the stellar surface density towards Sgr\,A*.

While the evidence for a stellar cusp is now strong, its exact
properties remain elusive. All tracer stars (even the ones creating
the diffuse, unresolved light) are confined to narrow
mass intervals.  With BaSTI models and at the distance and
  extinction (assuming $A_{K}=2.5$\,mag) of the GC all stars of
  brightness $19\lesssim K<10$ are confined to the narrow mass interval
  of $1.05-1.10$\,$M_{\odot}$ at 11\,Gyr age and to
  $1.45-1.55$\,$M_{\odot}$ at 3\,Gyr age
  \citep[see also][]{Baumgardt:2018ad,Schodel:2018db} and should therefore show
  the profile of a classical Bahcall-Wolf cusp with a power-law
  exponent of $\gamma=-1.5$ for the 3D stellar density
  \citep{Bahcall:1976vn,Alexander:2017fk}. The projected densities
  found here and in the above cited publications are mostly flatter
  than what would be expected for this value.  In particular when the
  2D data are de-projected, then relatively the power-law exponent
  takes on relatively small values \citep[as low as $\gamma=-1.1$,
  see][]{Schodel:2018db}.

Apart from possible systematic effects, in particular when
de-projecting the data, a decisively limiting factor in our analyses
is the unknown age of individual stars. \citet{Aharon:2015ij} and
\citet{Baumgardt:2018ad} have shown how repeated star formation can
create a cusp that is flatter than the classical Bahcall-Wolf cusp,
which is based on the simplifying assumptions of a single-age,
single-mass, old, dynamically relaxed population.  To properly
investigate the shape of the cusp, we would need to know the age of
the tracer stars. This requires careful spectro-photometric
work. Also, currently, efficient spectroscopy is limited to giants
brighter than $K_{s}\approx16$ and to small fields due to the small
fields-of-view of adaptive optics assisted integral field
spectrographs. The work of \citet{Habibi:2019tl} presents a first step
into this direction, but a more global and detailed study of all
spectro-photometric data (including a complex completeness analysis)
is necessary to address this problem. Alternatively, we may wait for
the advent of Extremely-Large-Telescopes to solve this question.

\begin{figure}[!htb]
\includegraphics[width=\columnwidth]{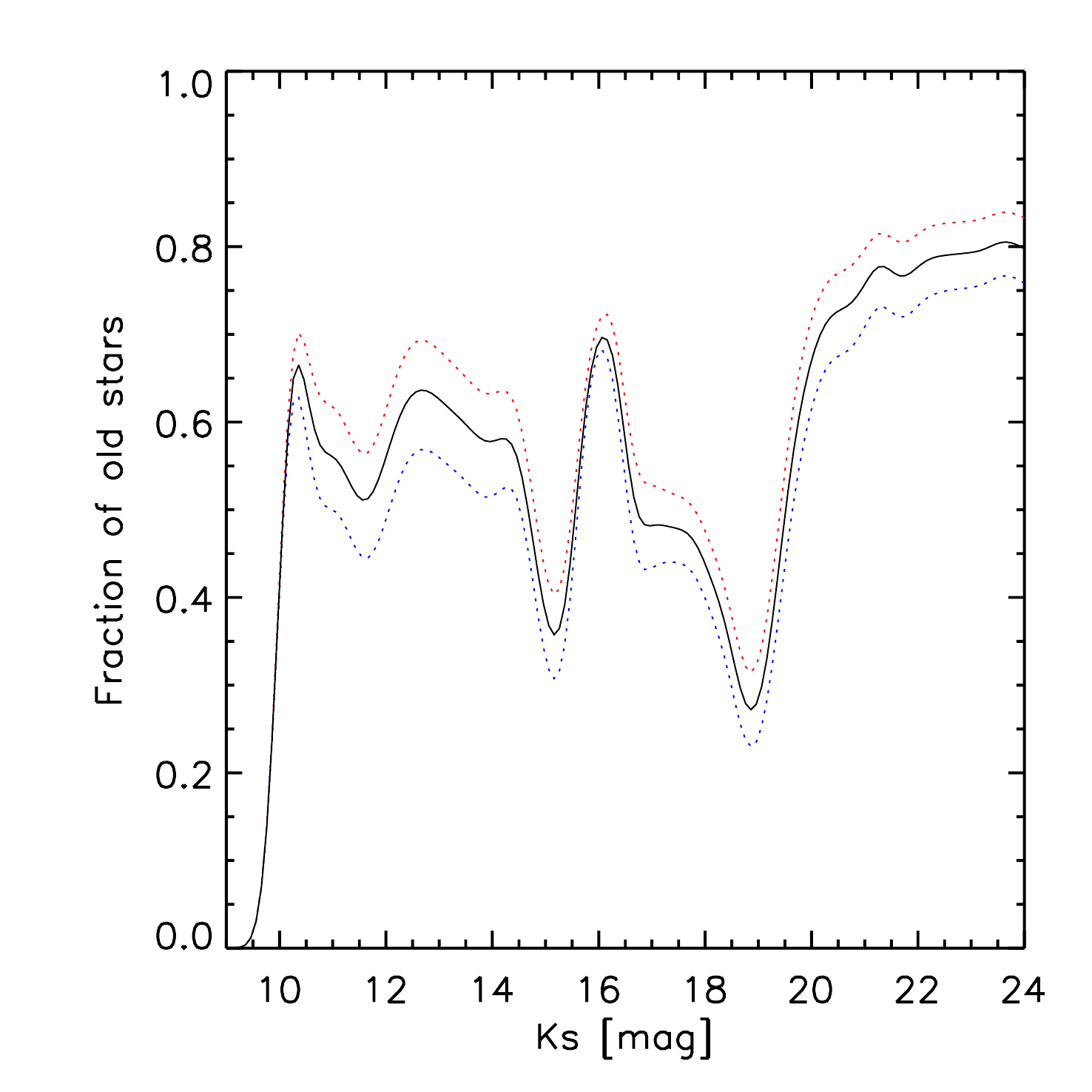}
\caption{\label{Fig:f_old} Fraction of observed stars older than
  ~10\,Gyr as a function of magnitude at the GC, based on the MC
  modelling of the data with BaSTI models and mean $[Fe/H$ of $1.5$
  solar (corresponding to the SFH shown in
  Fig.\,\ref{Fig:SFH_MC}). The lines correspond to cumulative
  probabilities of the simulations: 5\% or less of the MC runs (blue
  dotted), 50\% or less of the MC runs (black solid), 95\% or less of
  the MC runs (red dotted).}
\end{figure}

\citet{Gallego-Cano:2018nx} presented a plot of the observable
fraction of stars at a given magnitude that is older than 3\,Gyr
(their Fig.\,11). Here, we present a similar plot, based on the
  MC simulation from section\,\ref{sec:MC}. In Fig.\,\ref{Fig:f_old}
  we show the fraction of stars older than 10\,Gyr, along with
  confidence intervales. As concerns a comparison of this plot with
  the one shown in \citet{Gallego-Cano:2018nx}, the fraction of old
  stars is lower because the latter was based on the SFH derived by
  \citet{Pfuhl:2011uq} and considered stars older than 3\,Gyr. In our
  newly derived SFH we find no relevant star formation at ages
  5-9\,Gyr. This implies an overall higher contamination by young,
  dynamically unrelaxed stars at magnitudes brighter than
  $K\approx20$.

We can see the fraction of old, dynamically relaxed stars among
  the currently observable ones is only around 60\% at most
  magnitudes. Potential contamination by young stars is lowest at the
  peak of the RC, at $K\approx16$, and highest at $K\approx19$.
  Consequently, it is not straightforward to compare the observations
  to the theoretical predictions, in particular quantitatively. With
  next generation AO instruments and/or extremely large telescopes we
  will be able to probe the old main sequence stars at $K\gtrsim20$
  and thus derive a much clearer picture of the cusp around Sgr\,A*.

\subsection{Stellar remnants}

Pulsars in the GC are of great interest because they can serve as
exquisite probes of General Relativity if they are located in a tight
orbit around Sgr\,A* or, alternatively, form a binary with a stellar
mass black hole \citep[e.g.\ ][]{Eatough:2015fm}.  We find that 
  not more than about 1\% of the NSC may have formed less than
  100\,Myr ago. Stars with masses between about 8 to 25\,$M_{\odot}$
that formed during this time may have given rise to still active
pulsars (assuming a 100\,Myr {\it death line} due to spin-down). 
  As a very rough estimate, assuming conservatively that all these
  young stars formed within 1\,pc of Sgr\,A* and that there are
  $~1\times10^{6}$\, $M_{\odot}$ of stellar mass within 1\,pc of
  Sgr\,A*, this results in
  $\sim$$100$ pulsars, assuming a Salpeter IMF and a lower mass
  cut-off of
  0.5\,$M_{\odot}$. This number is not very sensitive to changes in
  the slope of the IMF: While a more top heavy IMF will produce more
  massive stars, it will also produce less stars in the mass interval
  where they end up in neutron stars. However, a lower mass cut-off of
  the IMF can change this number. For example, for a lower mass cut
  off of 1
  (2)\,$M_{\odot}$, the number of expected pulsars increases by a
  factor of 3 (7).

Of the order 20\% of these pulsars could be beamed towards Earth
\citep{Eatough:2015fm}. Indeed, a magnetar was detected at 0.1\,pc
projected distance from Sgr\,A* \citep{Rea:2013fk}. However, given the
small number of expected pulsars, it is highly unlikely that any
normal pulsar may be detected that is on a tight orbit around Sgr\,A*
and is beamed towards Earth. Previous estimates of up to 1000 pulsars
on short-period
($<100$\,yr) orbits around Sgr\,A* appear to be significant
overestimations \citep{Pfahl:2004rx}. The search for so-called
'recycled' pulsars, milli-second pulsars spun up by accretion onto old
neutron stars, may be more promising \citep{Eatough:2015fm}.

As concerns the predicted cusp of stellar mass black holes around
Sgr\,A*, stellar mass black holes constitute the heaviest long-lived
constituents of the stellar cluster they will mass-segregate towards
the central black hole and form a steep cusp around it, which may also
be termed {\it dark cluster} \citep[see][and references
therein]{Alexander:2017fk}. The existence of a stellar cusp around
Sgr\,A* suggests that stellar cusps and thus dark clusters exist
around similar massive black holes in the Universe. In this case,
future space-borne gravitational wave observatories will probably
detect a significant rate of so-called extreme mass ratio inspiral
events \citep[EMRIs, see][and references
therein]{Amaro-Seoane:2007ve}. In simulations, stellar mass black
holes are typically assigned masses of 10\,$M_{\odot}$. The higher the
metallicity of a star, the more mass it looses via stellar winds
during its post-main sequence evolution. This will result in smaller
remnant masses. The stellar mass black holes formed in the
high-metallicity NSC may therefore have preferentially of the order
6\,$M_{\odot}$ for solar metallicity, and even less for supersolar
metallicity precursors \citep[e.g.\ ][]{Banerjee:2019qo}. This may
slow down mass segregation and result in a somewhat less steep cusp of
the dark cluster.

\subsection{Formation of the MWNSC}

The two basic formation scenarios suggested for NSCs are
globular cluster infall and merger or in-situ growth by star formation
within the cluster or accretion of star clusters formed in its
vicinity \citep{Boker:2010ys,Neumayer:2017fk}. In case of the Milky
Way, there is direct evidence for the in-situ growth scenario
from the observation of young, massive stars close to Sgr\,A*
\citep[see review by][]{Genzel:2010fk}.

If globular clusters were the precursors of a significant fraction of
the MWNSC's mass, we would expect that a very old, sub-solar
metallicity population made up a large fraction of its mass.  The fact
that no RR Lyrae stars have been discovered in the central few parsecs
(or at most a single one) points towards a very small fraction of such
an old, metal poor population \citep{Dong:2017zl}. This agrees well
with the spectroscopic observations by \citet{Do:2015ve} and
\citet{Feldmeier-Krause:2017kq,Feldmeier-Krause:2020pi}, who find only
of the order 5\% of stars with metallicity of half solar or
less. Recent spectroscopic studies \citep{Schultheis:2019lw} and the
photometric analysis in this work support a supersolar mean
metallicity. The catalogue by \citet{Harris:2010zl} lists 152
  Milky Way globular clusters with valid metallicity
  measurements. There are only five globular clusters with a
  metallicity $>-0.3$\,dex and 97\% of the clusters in the sample have
  a lower metallicity.  All these observations therefore 
  disfavour globular clusters, such as they are observed today in the
  Milky Way, as MWNSC precursors.

The new SFH of the MWNSC derived in this work appears to push back its
original formation to a time $\sim$10\,Gyr ago, when the Milky Way may
have suffered its last major mergers
\citep{Wyse:2001bx,Helmi:2018sy}. Such mergers are efficient in
funnelling gas to galaxy centres and can thus lead to growth of
nuclear star clusters and massive black holes. Dense, metal rich star
clusters may have formed closed to the Galactic centre and then have
merged to form the MWNSC.  By forming originally in dense clusters close
to the centre but not in accretion discs around the central black
hole, the IMFs of these cluster may have avoided to be strongly
top-heavy or bottom-truncated IMF as it has been observed for the
youngest stars within 0.5\,pc of Sgr\,A*, which may have formed in
such an accretion disc
\citep[see][]{Bartko:2010fk,Genzel:2010fk,Lu:2013fk}. The IMF of the
NSC can therefore be closed to standard, as is also suggested by
observational work \citep{Lockmann:2010fk}, even though the newest
batch of stars has a top-heavy IMF. 

\subsection{Comparison to SFH of nuclear disc}

The SFH of the nuclear stellar disc, that surrounds the MWNSC, has
recently been published by \citet{Nogueras-Lara:2019qd}. They find
that 80\%-90\% of the stellar mass in the NSD formed over 8\,Gyr ago,
followed by a long minimum in activity that was ended by a starburst
like event about 1\,Gyr ago. They also find increased star formation
activity in the past few 100 Myr. The general old age and the high
activity in the recent past are consistent with the star formation
history of the MWNSC as derived here. There are two notable
differences: (1) \citet{Nogueras-Lara:2019qd} do not report any star
forming activity in the nuclear disc at intermediate ages, contrary to
the very clear signs of an about 3\,Gyr old population that we find in
the MWNSC. Either the latter event was limited by some mechanisms to the
MWNSC or it was missed in the study of \citet{Nogueras-Lara:2019qd}
because their data are of lower angular resolution and therefore
limited by crowding to magnitudes $\lesssim17$. The intermediate age
population manifests itself in the form of an upturn of the KLF at
$K>17.5$ (see Fig.\,\ref{Fig:SFH_BEST}). (2) The star burst event
about 1\,Gyr ago found by \citet{Nogueras-Lara:2019qd} in the NSD does
not have any counterpart in the MWNSC. It may have been limited to the
NSD with too little gas reaching the MWNSC.

\section{Summary and conclusions}

We present a new photometric analysis of the Milky Way's nuclear star
cluster in a region within about 1\,pc in projection around the
massive black hole Sgr\,A*. The stacked high-quality images result in
a star list for what we call the {\it large field} here, containing
$39,000$ $K_{s}$ magnitudes, which is complemented by $11,000$ $H$
measurements. The smaller number of $H$-band data points is caused by
the significantly higher extinction in $H$ and by the availability of
less high quality images in that band. In addition, we provide a deep
image of a small region of about $10"\times10"$ that results from
stacking the AO images with the speckle holography technique. The
latter leads to a narrow, Gaussian PSF across the entire image and
compensates the relative changes of the PSF across the individual
exposures. Applying the speckle holography technique to the AO data
significantly reduces incompleteness due to crowding and results in
about 50\% more stars detected at faint magnitudes as compared to the
classical shift-and-add technique. We also provide the source list of
the $\sim$$3,000$ stars detected at $K_{s}$ in this so-called {\it deep
 field}.

An analysis of the sensitive, extinction and completeness corrected
KLF obtained from our data shows that there is no detectable
dependence of the KLF on distance from Sgr\,A*. The only exception is
a slightly increased number of stars at $K_{s}=12-14$ within a
projected distance of 0.5\,pc from Sgr\,A*, which is due to the young,
massive stars present in this region. The super-exponential increase
of the KLF at faint magnitudes is evidence for the presence of
intermediate age stars (2-4\,Gyr), while its high level on the faint
side of the Red Clump bump provides evidence for the presence of young
(less than a few 100\,Myr) stars.

We derive new constraints on the star formation history (SFH) of
  the NSC from the KLF. A large fraction ($\sim80\%$) of the stars
  formed $\gtrsim$10\,Gyr ago, followed by almost zero activity during
  more than 5\,Gyr. Significant star formation occurred then again
  $\sim$3\,Gyr ago, when roughly 15\% of the original stellar mass were
  formed. There was a clear minimum in star formation around
  0.8-1\,Gyr ago and a few percent of the mass formed in the past few
  100 Myr. This is not consistent with a quasi-continuous SFH as
  hypothesised by \citet{Morris:1996vn}.  The spectroscopic and
  photometric data can be fit with a simple model, where most of
  the MWNSC formed $\sim$10\,Gyr ago, with additional star
  formation at $\sim$3\,Gyr and in the recent past (100\,Myr ago to
  present).

We use the data from the central holographic image to probe for the existence of the
predicted stellar cusp at the faintest magnitudes studied via star
counts yet. Consistent with previous work we find evidence for a
power-law increase of the stellar surface density at all magnitudes
fainter than $K_{s}\approx15$. Brighter giants show a flat profile in
the inner
$\sim$$5''$/$0.2$\,pc. With the help of our analysis of the SFH
history we show that the fraction of dynamically relaxed stars  (here
assumed to be older than about 10\,Gyr), which are adequate tracers of the
cusp, appears not be much higher than 60\% at the observed magnitudes. This means
that contamination of the star counts by younger, dynamically
unrelaxed stars can be significant at all magnitudes. While the
detection of similar power-law profiles at all magnitude bins and even
in the faint, unresolved light strongly support the existence of a
stellar cusp, the potentially high contamination makes it difficult to
infer its properties accurately.

From the SFH we argue that the detection of normal pulsars in tight ($<
100$ year period) orbits around Sgr\,A* is unlikely. The high mean
metallicity of the MWNSC may imply stellar black hole masses
significantly smaller than the typically assumed 10\,$M_{\odot}$. 

The high mean metallicity of the MWNSC and its star formation history,
which may be quasi-continuous, argue against the hypothesis that
globular clusters  -- such as they are observed in the Milky Way's
  halo -- have contributed a significant amount of the mass of
the MWNSC. However, the MWNSC may have formed from metal-rich dense
clusters that formed near the Galactic Centre after gas infall in a
major merger.

Further progress in the field can be achieved by studying the stellar
density profile using a carefully calibrated sample of stars age-dated
via spectrophotometry. Improved instrumentation for AO near-infrared
imaging (perhaps the upcoming ERIS at the ESO VLT) may allow us to
probe deeper into the faint end of the KLF and thus to obtain better
constraints on the relative contributions from the oldest and
intermediate-old populations. Adding mid-infrared photometry from the
James-Webb-Space Telescope will finally allow us to break the
degeneracy between reddening and intrinsic stellar colours and use
colour-magnitude diagrams as diagnostics for the formation history of
the MWNSC. The superb capabilities of spectroscopic and
imaging instrumentation (MOSAIC, HARMONI) at the ESO ELT (Extremely
Large Telescope) will probably finally enable us to unambiguously infer
the properties of the stellar cusp around Sgr\,A* and to determine the
SFH of the NSC in detail.

%--------------------------------------------------------------------

\begin{acknowledgements}
  The research leading to these results has received funding from the
  European Research Council under the European Union's Seventh
  Framework Programme (FP7/2007-2013) / ERC grant agreement
  n$^{\circ}$ [614922]. RS, FNL, EGC, ATGC, and BS acknowledge
  financial support from the State Agency for Research of the Spanish
  MCIU through the "Center of Excellence Severo Ochoa" award for the
  Instituto de Astrof\'isica de Andaluc\'ia (SEV-2017-0709). ATGC, BS,
  and RS acknowledge financial support from national project
  PGC2018-095049-B-C21 (MCIU/AEI/FEDER, UE).  F. N.-L. gratefully
  acknowledges funding by the Deutsche Forschungsgemeinschaft (DFG,
  German Research Foundation) -- Project-ID 138713538 -- SFB 881
  (``The Milky Way System'', subproject B8). This work is based on
  observations made with ESO Telescopes at the La Silla Paranal
  Observatory under programmes IDs 083.B-0390, 183.B-0100 and
  089.B-0162. We thank the staff of ESO for their great efforts and
  helpfulness.
\end{acknowledgements}

\bibliography{/Users/rainer/Documents/BibDesk/BibGC}

\appendix

\section{Systematic uncertainties in the KLF}
\label{app:KLF_systematics}

\begin{figure}[!htb]
\includegraphics[width=\columnwidth]{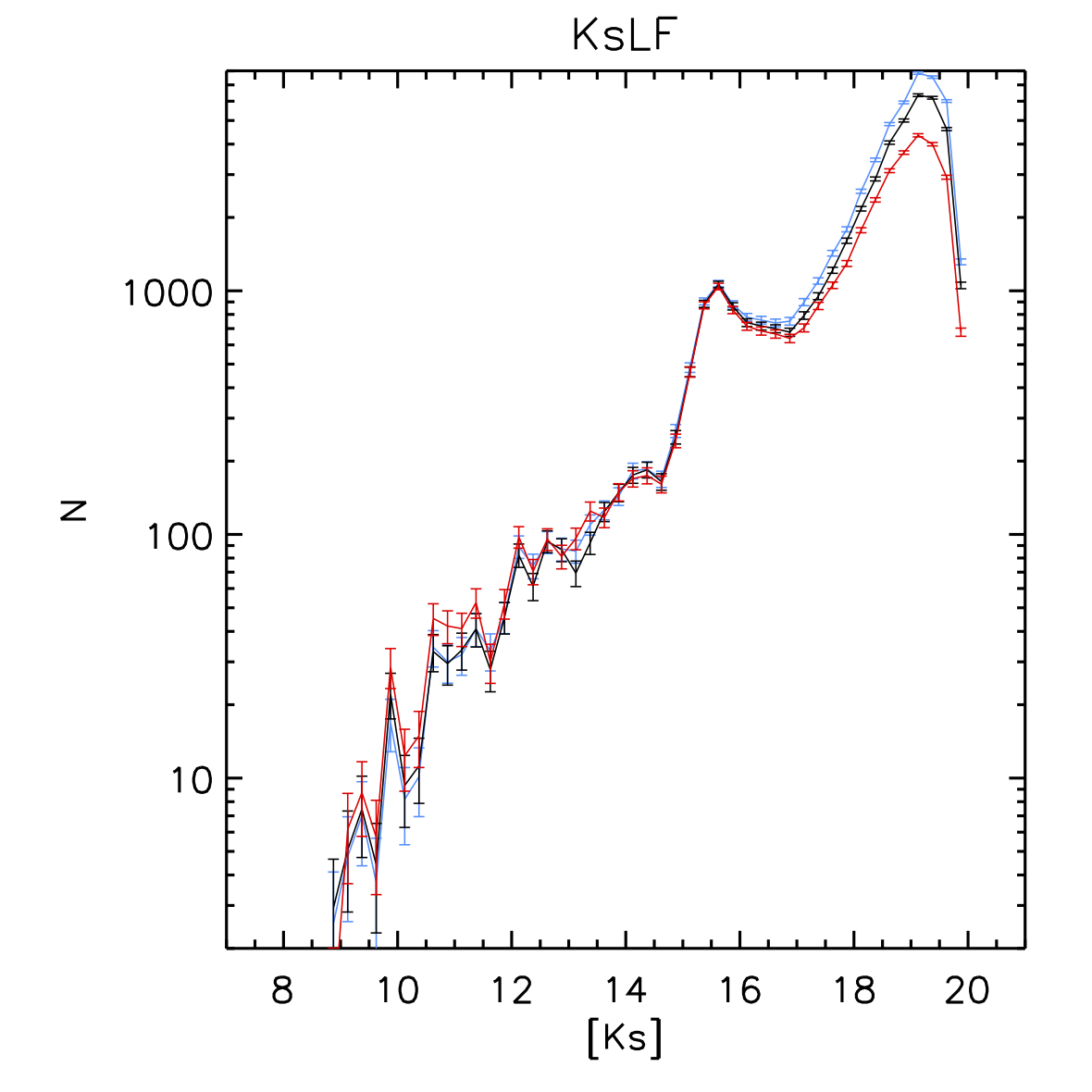}
\caption{\label{Fig:correlation} KLFs derived from the $K_{s}$ mosaic
  with $\starf$, using correlation thresholds of $0.6$ (blue line),
  $0.7$ (black), and $0.9$ (red).}
\end{figure}

The choices of parameters for source extraction and completeness
correction may cause systematic effects in the measured KLFs. The
correlation threshold set in $\starf$ to decide whether a source is a
star or not has a direct effect on the number of sources detected in
an image. If it is very low, many spurious sources may be picked
up. If it is too high, we will miss many real sources. A typical
choice of the correlation value is $0.7$, which is also the default
setting of $\starf$. We explored using values of $0.6, 0.7$, and $0.9$
on the $K_{s}$ mosaic. Figure\,\ref{Fig:correlation} shows the
resulting KLFs. The star counts start diverging significantly at
$K_{s}\gtrsim18$. In the main text we use a correlation threshold of
$0.7$, which is also the value used for the sources reported in
Table\,\ref{Tab:list}. We determine the standard deviation of the
three measurements for the different correlation thresholds and add it
quadratically to the uncertainties of the KLF for correlation
threshold $0.7$, which is used in the analysis in the main text.

\section{Additional results for MC simulation}
\label{app:MC}

\begin{figure}[!htb]
\includegraphics[width=.95\columnwidth]{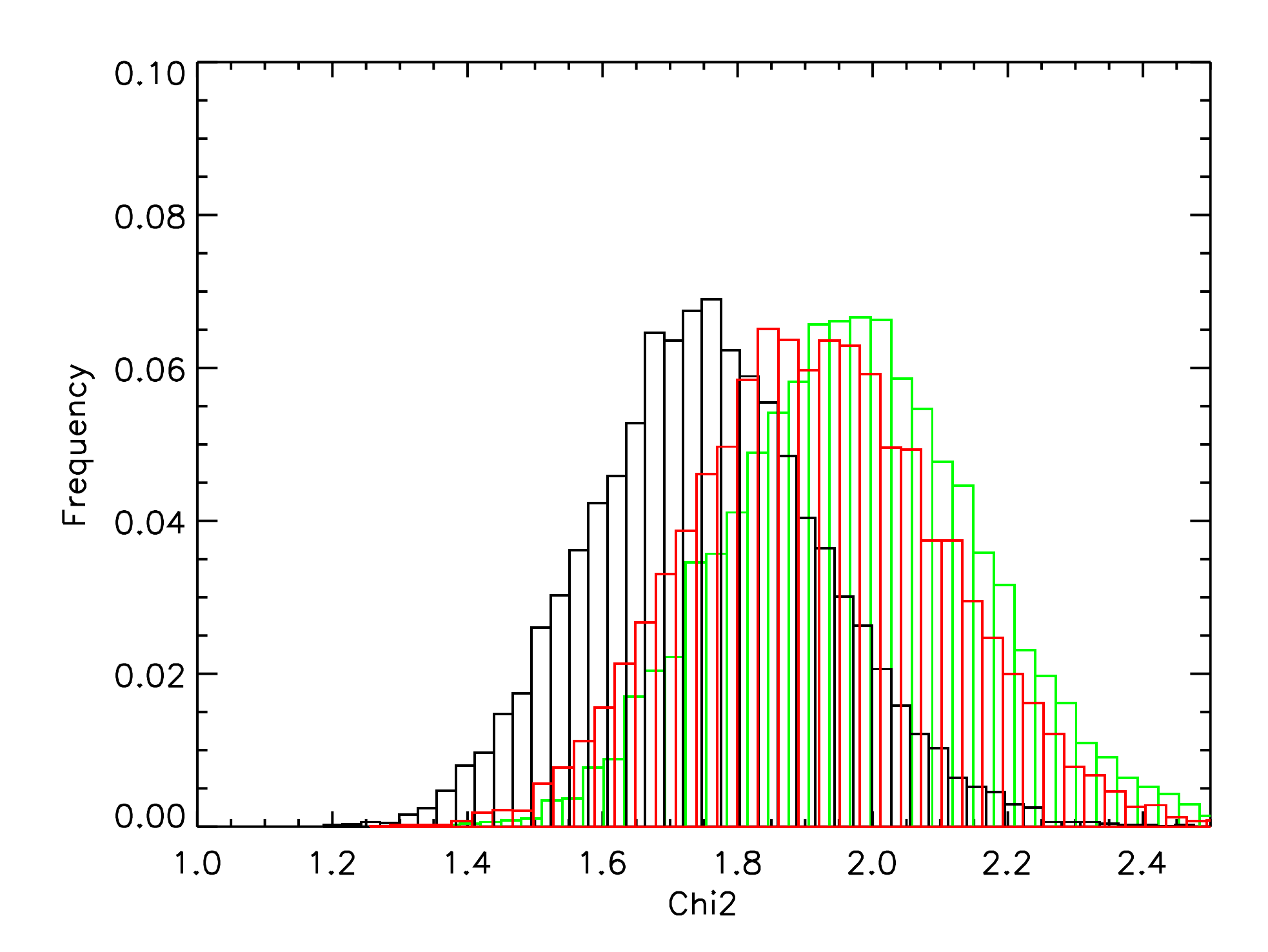}
\caption{\label{Fig:MC_chi2s} Best fit reduced $chi^{2}$ values for
  the fits to the MC simulated data for mean $[Fe/H]$ of solar
  (green), $1.5$ solar (black), and $2.0$ solar (red).}
\end{figure}

\begin{figure}[!htb]
\includegraphics[width=.95\columnwidth]{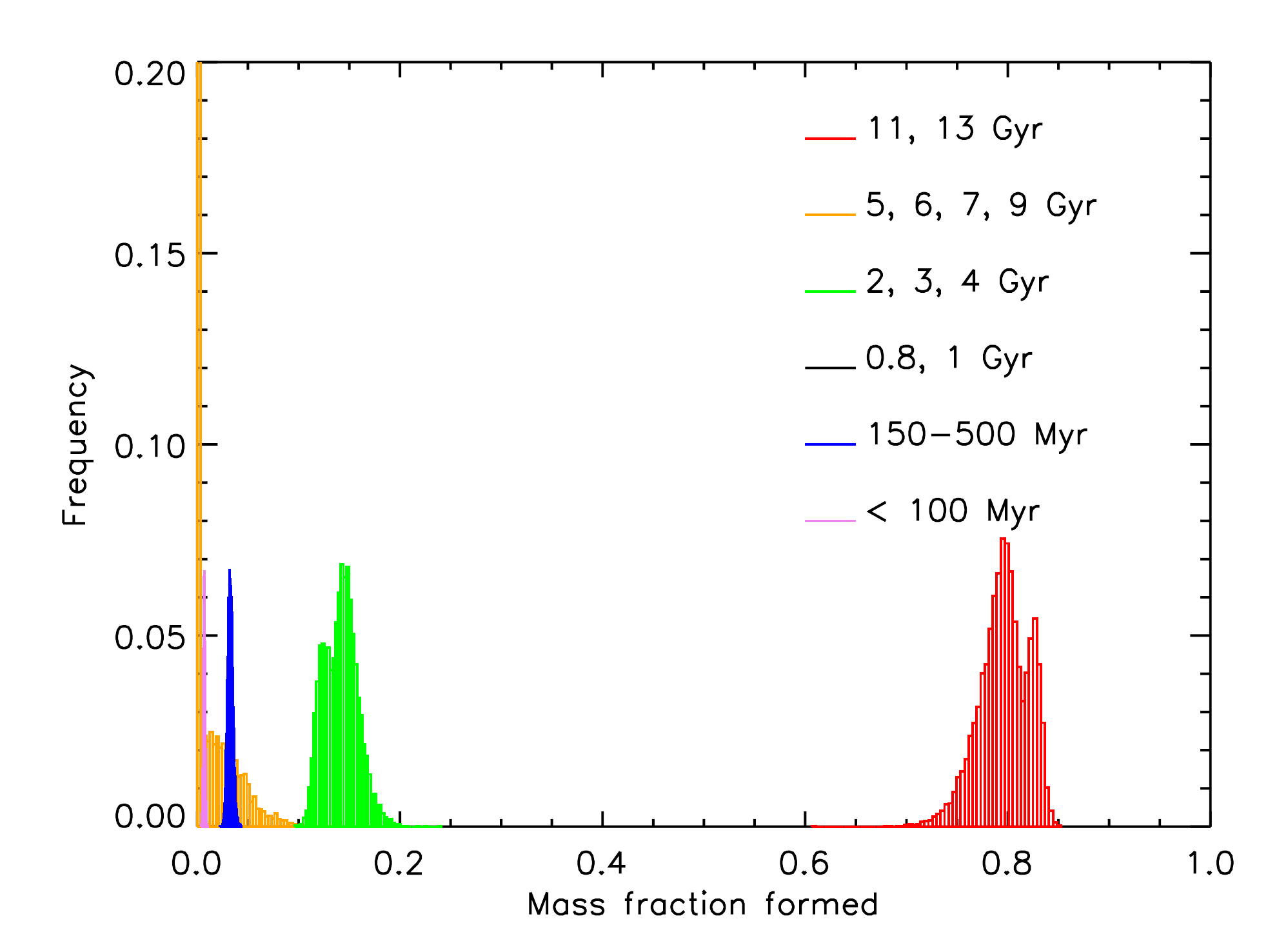}
\caption{\label{Fig:SFH_MC_BASTI1Z} Histograms of the stellar mass fraction
  formed in given age ranges derived from fits to MC simulations of
  the cumulative KLF with BaSTI isochrones and assuming a mean $[Fe/H]$
  of $1.0$ times solar.}
\end{figure}

\begin{figure}[!htb]
\includegraphics[width=.95\columnwidth]{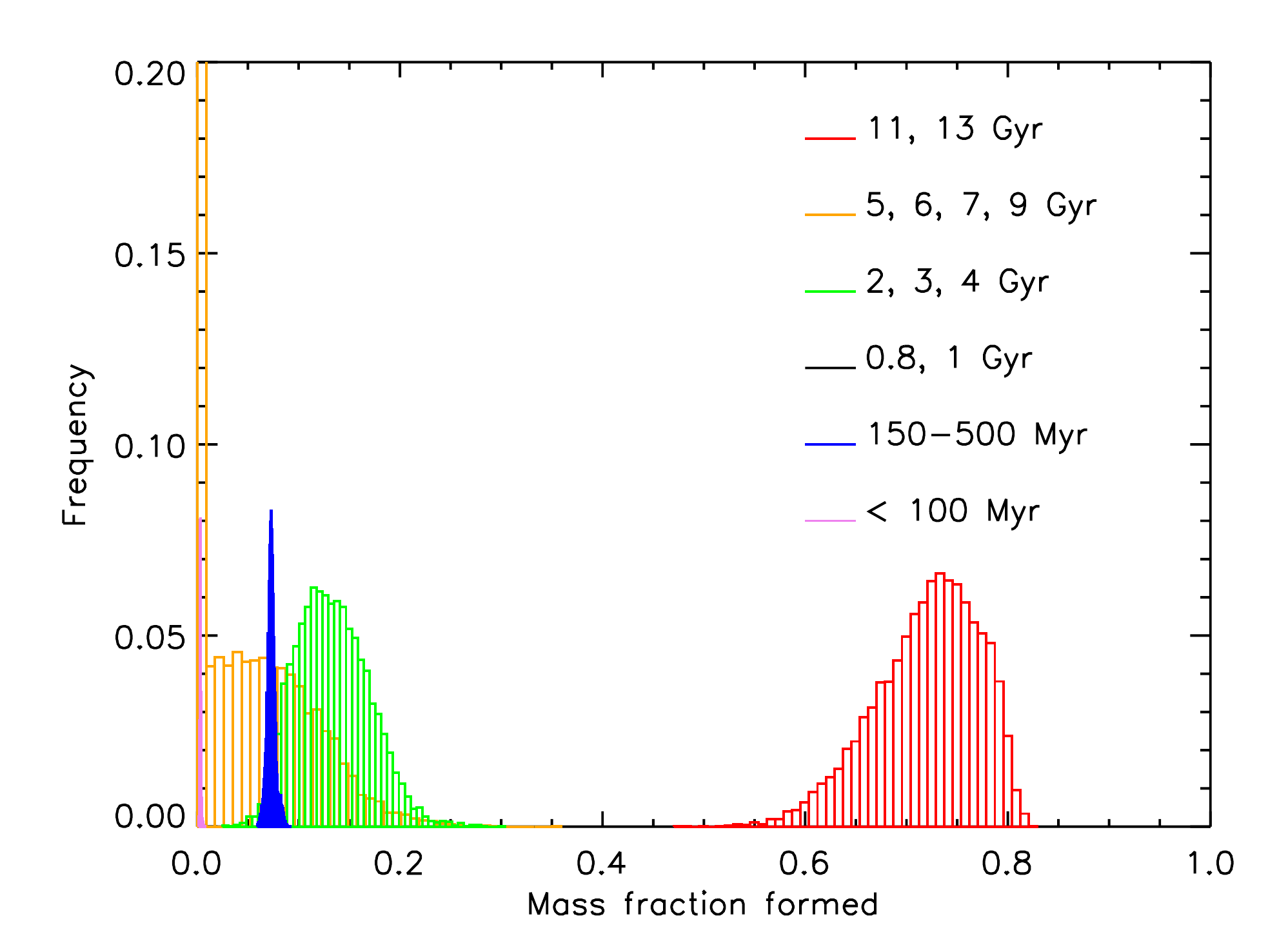}
\caption{\label{Fig:SFH_MC_MIST} Histograms of the stellar mass fraction
  formed in given age ranges derived from fits to MC simulations of
  the cumulative KLF with MIST isochrones and assuming a mean $[Fe/H]$
  of $1.5$ times solar.}
\end{figure}

\section{Tables of detected sources}

\begin{table*}[!htb]
  \caption{\label{Tab:list} List of detected point sources in the
    NACO $Ks$ and $H$-band large field mosaics.  We only include the
    first ten lines of the table in the printed edition of
    this work. A value $>99$ in the table indicates that  the corresponding measurement is not available.}
 \label{Tab:list_mosaic}
\centering
\begin{tabular}{rrrrrrrrr}
  \hline \hline
  R.A.\tablefootmark{a} &  Dec.\tablefootmark{b}  &  x & y &  $K_{s}$\tablefootmark{c}  &  $\Delta  K_{s}$\tablefootmark{d} & $H$\tablefootmark{e} & $\Delta  H$\tablefootmark{f}  & young \tablefootmark{g}  \\
 &  & [pixel] & [pixel] & mag & &  & & \\
  \hline
\hline
   17:45:40.0422 &  -29:00:22.5957   &      1855.890 &        1939.880 &           7.775 &           0.027 &           9.722 &           0.029 &   0\\
   17:45:41.1800 &  -29:00:46.8150    &      735.726 &         167.445  &          8.958  &          0.027 &          10.885 &           0.029 &   0\\
   17:45:40.2610 &  -29:00:27.1686 &        1640.090 &        1605.210  &          9.187 &           0.027  &         11.078  &          0.029 &   1\\
   17:45:41.0410 &  -29:00:22.5504 &         888.853 &        1956.410 &           9.196 &           0.027  &         11.220  &          0.029 &   0\\
   17:45:39.3937 &  -29:00:14.5744 &        2491.110 &        2522.010 &           9.253 &           0.027 &          11.256 &           0.029 &   0\\
   17:45:40.4742 &  -29:00:34.4292 &        1427.760 &        1072.310 &           9.259 &           0.027 &          11.266 &           0.029 &   0\\
   17:45:39.7906 &  -29:00:35.0284 &        2088.830 &        1019.040 &           9.397 &           0.027  &         12.748 &           0.029 &   0\\
   17:45:40.1109 &  -29:00:36.2632 &        1777.920 &         932.658 &           9.699 &           0.027 &          11.949  &          0.029 &   0\\
   17:45:41.2051 &  -29:00:38.8751 &         716.755 &         751.979  &          9.712 &           0.027  &         11.900  &          0.029 &   0\\
   17:45:40.8397 &  -29:00:33.9616 &        1074.340 &        1110.860 &           9.825 &           0.027 &          12.586 &           0.029  &  0 \\
 \end{tabular}
\tablefoot{
\tablefoottext{a}{Right ascension.}
\tablefoottext{b}{Declination.}
\tablefoottext{c}{Magnitude in the NACO $K_{s}$ band filter.}
\tablefoottext{d}{Uncertainty in $K_{s}$. The $1\,\sigma$ uncertainty of the zero point, is$\Delta ZP_{K_{s}}=0.06$.}
\tablefoottext{e}{Magnitude in the NACO $H$ band filter.}
\tablefoottext{f}{Uncertainty in $H$. For more information see also footnote $^{\mathrm{e}}$. The $1\,\sigma$ uncertainty of the
  zero point is $\Delta ZP_{H}=0.06$.}
\tablefoottext{g}{Spectroscopically identified early type star,
  according to \citet{Do:2009tg} and \citet{Bartko:2009fq}.}
}
\end{table*}

\begin{table*}[!htb]
  \caption{\label{Tab:list_central} List of detected point sources in
    the deep, central NACO $Ks$ holographic image. The last column
    lists the fraction of bootstrap images, in which any given source
    was detected. A value $>99$ in the table indicates that the
    corresponding measurement is not available.}
 \label{Tab:list_mosaic}
\centering
\begin{tabular}{rrrrrrr}
  \hline \hline
  R.A.\tablefootmark{a} &  Dec.\tablefootmark{b}  &  x & y &  $K_{s}$\tablefootmark{c}  &  $\Delta  K_{s}$\tablefootmark{d} & detection frequency \tablefootmark{e} \\
 &  & [pixel] & [pixel] &mag  & & \\
  \hline
\hline
   17:45:40.2592 &  -29:00:27.1576     &    149.575    &     428.095     &      9.152  &   0.007  &      1.00\\
   17:45:40.1169  & -29:00:27.5128     &    288.144    &     399.977     &      9.835  & 0.004    &      1.00\\
   17:45:40.1242  & -29:00:29.0311     &    278.870    &     288.077     &     10.000  &   0.008  &     1.00\\
   17:45:39.9217  & -29:00:26.7214     &    478.121    &     455.703     &     10.039  &   0.003  &     1.00\\
   17:45:40.0470  & -29:00:26.8603     &    357.112    &     447.165     &     10.183    &  0.004 &     1.00\\
   17:45:40.0935  & -29:00:31.2332     &    307.259    &     125.243     &     10.196    &  0.005  &    1.00\\
   17:45:40.1893  & -29:00:27.4726     &    217.530    &     403.918     &     10.561    &  0.006  &    1.00\\
   17:45:40.1418  & -29:00:29.9707     &    260.729    &     219.002     &     10.656    &  0.005  &    1.00\\
   17:45:39.8100  & -29:00:29.7463     &    582.408    &     231.098     &     10.713    &  0.005  &    1.00\\
   17:45:39.7921  & -29:00:29.8077     &    599.586    &     226.320     &     10.713    &  0.005  &     0.99\\
 \end{tabular}
\tablefoot{
\tablefoottext{a}{Right ascension.}
\tablefoottext{b}{Declination.}
\tablefoottext{c}{Magnitude in the NACO $K_{s}$ band filter.}
\tablefoottext{d}{Uncertainty in $K_{s}$. The $1\,\sigma$ uncertainty of the zero point, is$\Delta ZP_{K_{s}}=0.06$.}
\tablefoottext{e}{Fraction of bootstrapped images in which the source
was detected.}
}
\end{table*}

\clearpage

\section{Cumulative KLF fits with different theoretical isochrones}
\label{sec:models}

\begin{table*}[!htb]
  \caption{Best model fits to the cumulative KLF, considering stellar
    populations of 16 different ages and different metallicities,
    using PARSEC isochrones. The columns after line 4, list the fraction of the
    total initially formed stellar mass corresponding to each age.}
 \label{Tab:PARSEC}
\centering
\begin{tabular}{llllllll}
  \hline 
  Age (Gyr) &  $0.5\,Z_{\odot}$ &  $1\,Z_{\odot}$  &  $1.5\,Z_{\odot}$  &  $2\,Z_{\odot}$   &  $1\,Z_{\odot}\tablefootmark{c}$ &  $1.5\,Z_{\odot}\tablefootmark{c}$ &  $2\,Z_{\odot}\tablefootmark{c}$ \\
  \hline
$\chi^{2}_{\mathrm red}$                              & $2.29$ & $0.95$ & $0.84$ & $0.90$ & $1.37$ & $1.21$ & $1.48$ \\ 
$<A_{K_{s}}>\tablefootmark{a}$                 & $2.76$ & $2.72$ & $2.75$ & $2.71$ & $2.74$ & $2.76$ & $2.76$ \\
FWHM$_{smooth}\tablefootmark{b}$           & $3.71$ & $3.73$ & $3.68$  & $3.63$ & $3.68$ & $3.13$& $3.19$  \\
\hline
13     &    $0.011$          & $0.014$ & $0.524$   & $0.482$ & $0.000$ & $0.381$ & $0.412$ \\
11     &    $0.000$          & $0.408$ & $0.000$   & $0.000$ & $0.282$ & $0.000$ & $0.000$ \\
 9      &    $0.000$          & $0.000$ & $0.000$   & $0.000$ & $0.000$ & $0.000$ & $0.000$ \\
 7      &    $0.000$          & $0.000$ & $0.000$   & $0.000$ & $0.000$ & $0.000$ & $0.000$ \\
 6      &    $0.000$          & $0.000$ & $0.000$   & $0.000$ & $0.000$ & $0.000$ & $0.000$ \\
 5      &    $0.000$          & $0.000$ & $0.000$   & $0.213$ & $0.000$ & $0.351$ & $0.232$ \\
 4      &    $0.855$          & $0.429$ & $0.263$   & $0.095$ & $0.620$ & $0.052$ & $0.125$ \\
 3      &    $0.000$          & $0.081$ & $0.137$   & $0.142$ & $0.000$ & $0.121$ & $0.118$ \\
 2      &    $0.000$          & $0.000$ & $0.000$   & $0.000$ & $0.000$ & $0.000$ & $0.000$ \\
 1      &    $0.028$          & $0.000$ & $0.000$   & $0.000$ & $0.029$ & $0.026$ & $0.031$ \\
$0.8$   & $0.000$          & $0.000$ & $0.000$   & $0.000$ & $0.000$ & $0.000$ & $0.011$ \\
$0.5$   & $0.041$          & $0.030$ & $0.025$   & $0.023$ & $0.041$ & $0.039$ & $0.041$ \\
$0.25$ & $0.021$          & $0.015$ & $0.014$   & $0.016$ & $0.014$ & $0.014$ & $0.015$ \\
$0.15$ & $0.000$          & $0.007$ & $0.010$   & $0.009$ & $0.000$ & $0.004$ & $0.004$\\
$0.1$   & $0.000$          & $0.000$ & $0.008$   & $0.010$ & $0.000$ & $0.006$ & $0.007$ \\
$ 0.08$& $0.028$          & $0.006$ & $0.004$   & $0.004$ & $0.006$ & $0.000$ & $0.000$ \\
  $0.03$ & $0.015$        & $0.011$ & $0.006$   & $0.006$ & $0.008$ & $0.007$ & $0.004$ \\
\hline
\end{tabular}
\tablefoot{
   \tablefoottext{a}{Mean extinction..}
  \tablefoottext{b}{FWHM of Gaussian smoothing parameter in units of bins.}
 \tablefoottext{c}{Spectroscopically identified young stars excluded.}
}
\end{table*}

\begin{table*}[!htb]
  \caption{Best model fits to the cumulative KLF, considering stellar populations
    of 16 different ages and different metallicities, using MIST isochrones. The columns
    after line 4, list the fraction of the total initially formed
    stellar mass corresponding to each age.}
 \label{Tab:MIST}
\centering
\begin{tabular}{lllllllll}
  \hline 
  Age (Gyr) & $0.5\,Z_{\odot}$ & $1\,Z_{\odot}$ &  $1.5\,Z_{\odot}$ &  $2\,Z_{\odot}$ & $0.5\,Z_{\odot}\tablefootmark{c}$ & $1\,Z_{\odot}\tablefootmark{c}$ & $1.5\,Z_{\odot}\tablefootmark{c}$  & $2\,Z_{\odot}\tablefootmark{c}$  \\
  \hline
$\chi^{2}_{\mathrm red}$                     & $1.33$ & $1.24$ & $1.24$ & $1.61$ & $2.87$ & $1.42$ & $1.42$ & $1.68$ \\ 
$<A_{K_{s}}>\tablefootmark{a}$         & $2.70$ & $2.70$ & $2.70$ & $2.72$ & $2.75$ & $2.75$ & $2.75$ & $2.83$ \\
FWHM$_{smooth}\tablefootmark{b}$   & $2.16$ & $3.02$ & $3.00$ & $3.76$ & $2.11$ & $2.14$ & $2.21$ & $3.26$ \\
\hline
13     &    $0.637$  & $0.682$ & $0.694$ & $0.324$ & $0.458$ & $0.752$ & $0.762$ & $0.473$ \\
11     &    $0.000$  & $0.000$ & $0.000$ & $0.000$ & $0.000$ & $0.002$ & $0.000$ & $0.021$ \\
 9      &    $0.000$  & $0.000$ & $0.000$ & $0.000$ & $0.000$ & $0.000$ & $0.000$ & $0.053$ \\
 7      &    $0.000$  & $0.000$ & $0.000$ & $0.000$ & $0.000$ & $0.000$ & $0.000$ & $0.000$ \\
 6      &    $0.000$  & $0.000$ & $0.000$ & $0.000$ & $0.000$ & $0.000$ & $0.000$ & $0.000$ \\
 5      &    $0.000$  & $0.376$ & $0.145$ & $0.567$ & $0.000$ & $0.000$ & $0.000$ & $0.000$ \\
 4      &    $0.098$  & $0.000$ & $0.000$ & $0.000$ & $0.321$ & $0.078$ & $0.066$ & $0.303$ \\
 3      &    $0.000$  & $0.131$ & $0.000$ & $0.000$ & $0.000$ & $0.000$ & $0.000$ & $0.000$ \\
 2      &    $0.155$  & $0.000$ & $0.107$ & $0.054$ & $0.017$ & $0.075$ & $0.080$ & $0.058$ \\
 1      &    $0.035$  & $0.024$ & $0.000$ & $0.000$ & $0.087$ & $0.019$ & $0.018$ & $0.007$ \\
$0.8$   & $0.000$  & $0.000$ & $0.000$ & $0.000$ & $0.000$ & $0.000$ & $0.000$ & $0.000$ \\
$0.5$   & $0.026$  & $0.038$ & $0.020$ & $0.013$ & $0.068$ & $0.053$ & $0.054$ & $0.060$ \\
$0.25$ & $0.019$  & $0.014$ & $0015$ & $0.021$  & $0.021$ & $0.013$ & $0.013$ & $0.016$ \\
$0.15$ & $0.016$  & $0.004$ & $0.010$ & $0.008$ & $0.016$ & $0.005$ & $0.005$ & $0.002$ \\
$0.1$   & $0.000$  & $0.006$ & $0.002$ & $0.009$ & $0.000$ & $0.000$ & $0.000$ & $0.004$ \\
$ 0.08$& $0.008$  & $0.000$ & $0.000$ & $0.000$ & $0.008$ & $0.000$ & $0.000$ & $0.000$ \\
$0.03$ & $0.006$  & $0.007$ & $0.006$ & $0.004$ & $0.004$ & $0.003$ & $0.003$ & $0.001$ \\
\hline
\end{tabular}
\tablefoot{
   \tablefoottext{a}{Mean extinction..}
  \tablefoottext{b}{FWHM of Gaussian smoothing parameter in units of bins.}
 \tablefoottext{c}{Spectroscopically identified young stars excluded.}
}
\end{table*}

\end{document}